\documentclass[aps,pra,twocolumn,superscriptaddress,nofootinbib]{revtex4-2}

\usepackage{graphicx,epsfig}
\usepackage{color}
\usepackage[usenames,dvipsnames]{xcolor}
\usepackage{amsmath,bbm,amssymb, amsthm}
\usepackage{dsfont} 
\usepackage{stmaryrd}
\usepackage{numprint}

\usepackage{hyperref}
\hypersetup{colorlinks=true,linktoc=all,linkcolor=blue,breaklinks=true,citecolor=blue,urlcolor=blue}

\def\bb{\begin{eqnarray}}
\def\ee{\end{eqnarray}}

\newcommand{\moy}[1]{\left\langle\!\!\!\left\langle #1 \right\rangle\!\!\!\right\rangle}

\begin{document}

\newcommand{\overbar}[1]{\mkern 1.5mu\overline{\mkern-1.5mu#1\mkern-1.5mu}\mkern 1.5mu}

\newcommand{\epsw}{\epsilon_\text{w}}
\newcommand{\epsh}{\epsilon_\text{h}}

\newcommand{\Gan}{\Gamma_{\alpha,n}}
\newcommand{\Ganbar}{\Gamma_{\alpha,n_{\overbar{j_\alpha}}}}
\newcommand{\GLn}{\Gamma_{\text{L},n}}
\newcommand{\GRn}{\Gamma_{\text{R},n}}
\newcommand{\GLN}{\Gamma_{\text{L},0}}
\newcommand{\GLO}{\Gamma_{\text{L},1}}
\newcommand{\GRN}{\Gamma_{\text{R},0}}
\newcommand{\GRO}{\Gamma_{\text{R},1}}
\newcommand{\Gwn}{\Gamma_{\text{W},n}}
\newcommand{\GwN}{\Gamma_{\text{W},0}}
\newcommand{\GwO}{\Gamma_{\text{W},1}}
\newcommand{\Gh}{\Gamma_{\text{H}}}

\newcommand{\nbar}{n_{\overbar{j_\alpha}}}
\newcommand{\nw}{n_\text{w}}
\newcommand{\nh}{n_\text{h}}
\newcommand{\dnw}{\dot{n}_\text{w}}
\newcommand{\dnh}{\dot{n}_\text{h}}
\newcommand{\Nw}{N_\text{w}}
\newcommand{\Nh}{N_\text{h}}
\newcommand{\dNw}{\dot{N}_\text{w}}
\newcommand{\dNh}{\dot{N}_\text{h}}
\newcommand{\C}{{\cal C}}
\newcommand{\Cb}{\bar{\cal C}}

\title{Stochastic Thermodynamic Cycles of a Mesoscopic Thermoelectric Engine}

\author{R. David Mayrhofer}
\affiliation{Department of Physics and Astronomy, University of Rochester, Rochester, NY 14627, USA}

\author{Cyril Elouard}
\email{cyril.elouard@gmail.com}
\affiliation{Department of Physics and Astronomy, University of Rochester, Rochester, NY 14627, USA}

\author{Janine Splettstoesser}
\affiliation{Department of Microtechnology and Nanoscience (MC2), Chalmers University of Technology, S-412 96 G\"oteborg, Sweden}

\author{Andrew N. Jordan}
\affiliation{Department of Physics and Astronomy, University of Rochester, Rochester, NY 14627, USA}
\affiliation{Institute for Quantum Studies, Chapman University, Orange, CA 92866, USA}

\date{\today}

\begin{abstract}
We analyze a steady-state thermoelectric engine, whose working substance consists of two capacitively coupled quantum dots. One dot is tunnel-coupled to a hot reservoir serving as a heat source, the other one to two electrically biased reservoirs at a colder temperature, such that work is extracted under the form of a steady-state current against the bias. In single realizations of the dynamics of this steady-state engine autonomous, 4-stroke cycles can be identified. The cycles are purely stochastic, in contrast to mechanical autonomous engines which exhibit self-oscillations. In particular, these cycles fluctuate in direction and duration, and occur in competition with other spurious cycles. Using a stochastic thermodynamic approach, we quantify the cycle fluctuations and relate them to the entropy produced during individual cycles. We identify the cycle mainly responsible for the engine performance and quantify its statistics with tools from graph theory. We show that such stochastic cycles are made possible because the work extraction mechanism is itself stochastic instead of the periodic time dependence in the working-substance Hamiltonian which can be found in conventional mechanical engines. Our investigation brings new perspectives about the connection between cyclic and steady-state engines.
\end{abstract}

\maketitle

\section{Introduction}

In recent years, there has been increasing interest in electronic systems working as heat engines at the nanoscale, see, e.g., Refs.~\cite{Sothmann14,Thierschmann2016Dec,Benenti2017Jun,Binder2018} for reviews.
One of the distinguishing advantages of devices at the nanoscale is that quantum confinement can be exploited for the implementation of heat engines. For example, in electronic heat engines using quantum dots, their discrete energy spectrum is exploited to increase the efficiency of heat-to-work conversion~\cite{Hicks1993May,Hicks1993Jun,Mahan1996Jul,Heremans2013Jun,Sothmann14} or to control the engine's operation mode. Furthermore, the smallness of the device ensures that thermal equilibrium is not reached during operation; only the reservoirs it is connected to are in an internal equilibrium state at well-defined temperatures and electrochemical potentials. This local nonequilibrium can serve as an additional source for power production~\cite{Sanchez2019Nov,Hajiloo2020Aug}.  In contrast to macroscopic engines, fluctuations are important when the system is small, often requiring a theoretical description in terms of stochastic thermodynamics~\cite{Esposito09,Seifert12}.

\begin{figure}[b]
\begin{center}
\includegraphics[width=\columnwidth]{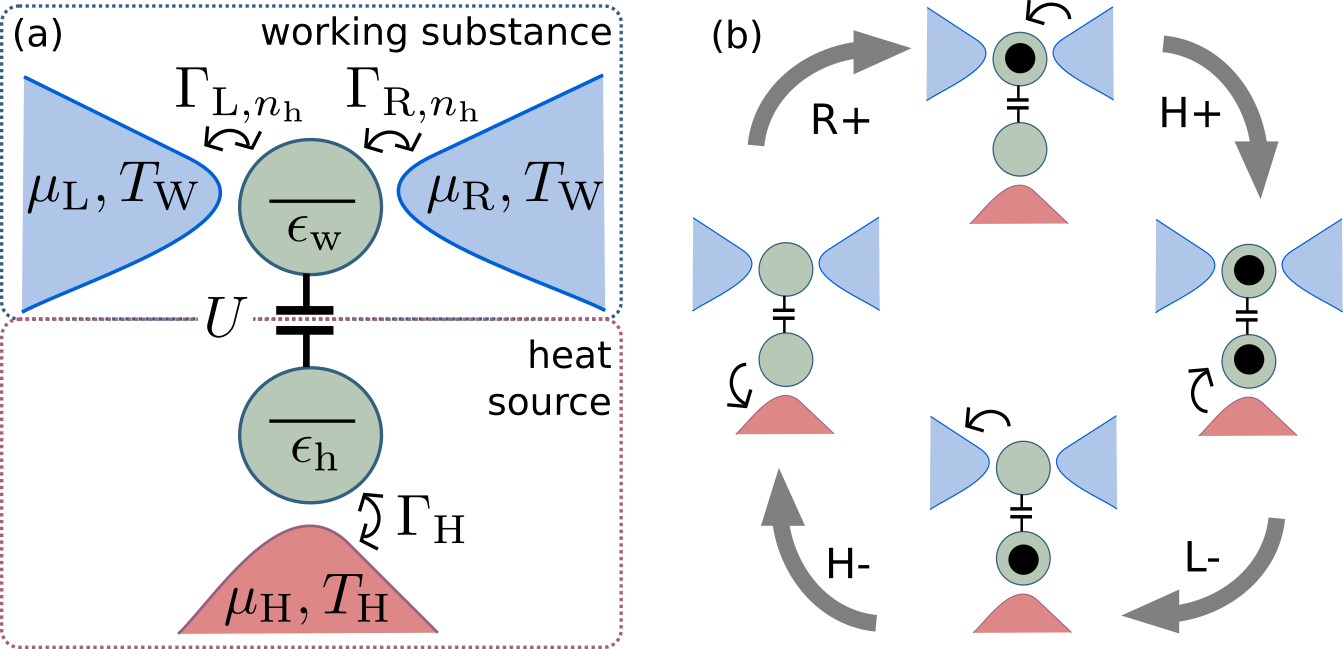}
\end{center}
\caption{(a) Sketch of the thermoelectric three-terminal engine that can be imagined as consisting of a heat source and a working substance. The double quantum dot contains a ``work" and a ``hot" dot with single-particle energies $\epsilon_\text{w}, \epsilon_\text{h}$, which are capacitively coupled with coupling energy $U$. The dots are coupled to reservoirs, characterized by electrochemical potentials $\mu_\text{L}$, $\mu_\text{R}$, and $\mu_\text{H}$ and temperatures $T_\text{W}$ and $T_\text{H}$. Importantly, the tunnel-coupling strengths, $\Gamma_{\text{L},n_\text{h}},\Gamma_{\text{R},n_\text{h}},\Gamma_\text{H}$ can depend on the occupation of the dot, which is not involved in the tunneling process. (b) Example of a cycle in state space $\{00,01,10,11\}$ leading to power production. Tunneling transitions are labeled by the involved reservoir (L,R,H) and increase (+) or decrease (-) of dot occupations. }\label{fig:Engine}
\end{figure} 

Some of these nanoscale engines have a similar operational principle as classical heat engines where work exchange is mediated by a cyclically varying degree of freedom---i.e., the equivalent of a piston or pump~\cite{Kosloff2013May}. The working substance of such engines is thereby successively coupled and uncoupled to baths at different temperatures~\cite{Geva1992Feb,Juergens2013Jun,Dare2016Jan,Bustos-Marun2019Aug,Pekola2019Aug}. These engines can be \emph{autonomous}, i.e. the strokes can happen without an external periodic driving, owing to self-oscillations~\cite{Alicki2016Jan,Alicki2017Mar,Wachtler2019Jul} of the piston. In contrast, many other nanoscale electronic heat engines are steady-state (or continuous) engines, meaning that they are operated in a (nonequilibrium) steady state, see Refs.~\cite{Sothmann14,Benenti2017Jun} for reviews. Steady-state electronic heat engines are mostly working based on a type of thermoelectric effect, as e.g. in recent experimental realizations of quantum dot heat engines~\cite{Svensson2013Oct,Josefsson2018Jul,Prete2019May,Josefsson2019Jun,Jaliel2019Sep,Harzheim2019}, and can be classified as autonomous as well.

From a \textit{technical} perspective, an equivalence between the nonequilibrium steady-state operation and stochastic pumps~\cite{Esposito2015May,Uzdin2015Sep,Raz2016May} has been established, useful for the quantitative analysis of both but also to develop improved schemes. In contrast, the \textit{conceptual} difference between steady-state and cyclic engines was highlighted: steady-state engines were sometimes referred to as particle-exchange engines in Ref.~\cite{Humphrey2005Oct}. Other papers even raised doubts whether steady-state engines are able at all to perform work in the absence of underlying self-oscillations~\cite{Alicki2016Jan,Alicki2017Mar}.

In this paper, we focus on one important class of steady-state heat engines, namely three-terminal energy harvesters~\cite{Sanchez11,Sothmann2012May,Mazza2014Aug,Hartmann2015Apr,Roche2015Apr,Thierschmann2015Aug,Walldorf2017Sep,Erdman2018Jul,Sanchez2019Oct,Sanchez2019Nov,Whitney2016Jan}, see Fig.~\ref{fig:Engine}, and analyze their operational principle with respect to cycles underlying the dynamics. 
The working principle of three-terminal energy harvesters is to some extent similar to previously mentioned quantum-dot based engines: energy-filtering is crucial for power production. An advantage of these setups however is that the heat source and working substance are spatially separated from each other, see  Fig.~\ref{fig:Engine} and its caption for details. In particular, they are typically not connected to each other through an electronic flow, but rather by capacitive coupling --- in our case between two quantum dots; some three-terminal steady-state heat engines have two electronic and a third non-electronic heat bath~\cite{Entin-Wohlman2010Sep,Arrachea2014Sep,Henriet15,Entin-Wohlman17}. 

For the analysis of cycles occurring in the steady-state operation of these devices, we take two different perspectives. On one hand, (i) we note that the dynamics of the occupation of the quantum dot connected to the hot reservoir in the heat resource induces an effective time dependence in the Hamiltonian for the upper dot as part of the working substance and its energy-dependent coupling to the left and right reservoirs. This time-dependent occupation could hence be imagined to take over the role of the piston of a standard cyclic engine. However, we prove in this paper that the dynamics of this degree of freedom is not oscillatory but rather stochastic. Even more, we can demonstrate that no oscillations are visible in correlation functions either, in contrast with the case of the thermoelectric engine based on Josephson junctions analyzed in Ref.~\cite{Verteletsky20}.

On the other hand, (ii) it has been shown previously that the autonomous steady-state dynamics of such system can be decomposed into stochastic cycles~\cite{Horowitz2014Jul,Lin2020May} without referring to the notion of an effectively time-dependent Hamiltonian. We analyze these cycles in detail and show that they are responsible for the performances of the engine while still being fully stochastic, with no underlying oscillatory behavior. Both these perspectives conclusively show that non-periodic cyclic engines are possible.

The paper is organized as follows. We introduce the explicit model for the three-terminal energy harvester and its working principle in Sec.~\ref{sec:model}. We then start the analysis within a master equation approach for an \textit{ensemble description} in Sec.~\ref{sec:average}, where we study the dynamics of the dot occupations, see Sec.~\ref{sec:avdyn} as well as of correlation functions between different currents, Sec.~\ref{sec:corr}. 
The master-equation based approach yields the charge and heat currents over ensemble averages  and even goes beyond the study of average engine performance. However, such methods give little information about the system during single realizations of the engine operation and may therefore hide a cyclic behavior during individual realizations. In Sec.~\ref{sec:stochastic}, we therefore use a stochastic description
for trajectories in state space. We identify a family of configuration-space cycles followed by the engine, alternatively in a stochastic order and analyze the properties of the most prevalent cycles driving the engine. Finally, in Sec.~\ref{sec:semi}, we study cycles of the working substance dynamics driven by the stochastic time dependence of the occupation of the dot in contact with the hot reservoir. Conclusions are presented in Sec.~\ref{sec:conclusions}.

\section{Model and engine operation}\label{sec:model}

The particular thermoelectric engine we study is a three-terminal architecture connected via a two quantum-dot system, see Fig.~\ref{fig:Engine}(a). It has first been introduced as a three-terminal energy harvester by S\'anchez and B\"uttiker~\cite{Sanchez11} and has subsequently been realized in experiments in the same~\cite{Thierschmann2015Aug} or similar type of settings~\cite{Hartmann2015Apr,Roche2015Apr}. 

In the upper part---consisting of a quantum dot in contact with two reservoirs---power is produced or, in other words, work is done; we refer to this entire part of the setup as the working substance and indicate it with a subscript ``W". Also the upper quantum dot, characterized by a single-level energy $\tilde{\epsilon}_\text{w}$, is referred to as the ``work" dot (subscript ``w"). The two contacts it is tunnel coupled to have the same temperature $T_\text{W}$ but different chemical potential $\mu_\text{L}>\mu_\text{R}$. 
From now on, we set $\mu_\text{R}$ as the reference energy and define $\epsw\equiv \tilde{\epsilon}_\text{w}-\mu_\text{R}$ and $\Delta\mu=\mu_\text{L}-\mu_\text{R}$. 
The work is extracted via the production of an electric current against that bias. Importantly, the tunnel-coupling strengths between the work dot and the left and right contact are energy dependent, $\Gamma_\alpha(E)$ with $\alpha\in\{\text{L},\text{R}\}$. Such an energy dependence can physically be implemented e.g. by quantum point contacts~\cite{Sothmann2012May,Yang2019Jul,Mavalankar2013Sep,Maradan2014Jun,Torresani2013Dec,Gasparinetti2012Jun}.

The lower part of the setup acts as the heat source, referred to by a subscript ``H". It consists of a second quantum dot, with single-level energy $\tilde{\epsilon}_\text{h}$, referred to as the ``hot" dot (subscript ``h"). This second dot is tunnnel coupled, with coupling strength $\Gamma_\text{H}$, to a third reservoir  characterized by a higher temperature $T_\text{H} > T_\text{W}$ and chemical potential $\mu_\text{H}$. We use the electrochemical potential $\mu_\text{H}$ as the reference energy for the lower part of the setup and define $\epsh=\tilde{\epsilon}_\text{h}-\mu_\text{H}$. The work dot and the hot dot are coupled to each other uniquely capacitively, meaning that no particle exchange takes place between the dots. The coupling strength between dots is quantified by the inter-dot Coulomb interaction strength $U$.

We restrict our analysis to weak reservoir-dot couplings $\Gamma_\alpha \ll T_\text{H},T_\text{W}$ and to the case in which the intra-dot Coulomb interaction, namely the onsite interaction \textit{within} each of the dots, is the largest energy scale, such that the probability of finding two electrons in either of the dots is negligible. 
In this configuration, there are four possible charge states $\{ab\}$ of the pair of quantum dots, where $a,b\in\{0,1\}$, $a$ corresponding to the occupation of the work dot and $b$ to the hot dot\footnote{Note, that we neglect the spin degree of freedom of electrons here. Taking into account the spin degree of freedom would modify relaxation rates, see e.g. Refs.~\cite{Splettstoesser2010Apr,Schulenborg2016Feb}. Including spin degeneracy would make the analysis of the present paper more cumbersome, but would not modify this specific effect qualitatively. ``Spinless" electrons can in an experiment be realized by the application of a strong magnetic field.}. The latter are the eigenstates of the free two-dot Hamiltonian, which reads:
\bb
H_\text{QD} = \epsw \hat{n}_\text{w} + \epsh \hat{n}_\text{h} + U\hat{n}_\text{w}\hat{n}_\text{h},
\ee
where we have introduced the electron number operators in dot w, $\hat{n}_\text{w}$, and in dot h, $\hat{n}_\text{h}$. The associated eigen-energies are given by $0$, $\epsh$, $\epsw$, $\epsh+\epsw+U$. We emphasize that this Hamiltonian is not time-dependent. However, when focusing on an effective Hamiltonian of the work dot alone, an effective time-dependence can be introduced, as we will show in Sec.~\ref{sec:semi}.

Here, due to the energy dependence of the corresponding tunnel couplings $\Gamma_\alpha(E)$, the rates at which the tunneling events occur between the left/right reservoir and the work dot depend on the state of the hot dot. Instead of explicitly writing the energy dependence of $\Gamma_\text{L,R}(E)$, we  write $\Gamma_{\alpha,n_\text{h}}$ for the coupling rate to reservoir $\alpha=\text{L,R}$ when the hot dot occupation is $n_\text{h}$. This feature is essential for the setup to work as an engine~\cite{Sanchez11}. In the ideal case, the asymmetry
\begin{equation}\label{eq:asym}
    {\cal A} = \frac{\GRN\GLO-\GRO\GLN}{(\GLN+\GRN)(\GLO+\GRO)}\rightarrow 1
\end{equation} 
 is large\footnote{In the present paper, we impose one specific sign for the asymmetry, without however restricting generality. Ideal heat-to-work conversion is also realized in the limit of the asymmetry going to $-1$, if $\mu_\text{R}>\mu_\text{L}$.}. Note that a possible energy dependence of $\Gamma_\text{H}$, which would result in a dependence on the occupation of the work dot $n_\text{w}$, is not relevant for the working principle of the device and we therefore neglect it here, $\Gamma_\text{H}(E)\equiv\Gamma_{\text{H},n_\text{w}}\equiv\Gamma_\text{H}$.

When the asymmetry defined above is greater than zero---namely, $\GRN\GLO>\GRO\GLN$---the mechanism allowing for work extraction is the following: the work dot can exchange electrons with both reservoirs L and R, with a statistical preference of electric current flowing along decreasing electrochemical potentials, i.e. here from the left to the right reservoir. When the hot dot is empty, there is possibly a higher probability for tunneling events with the right reservoir. Conversely, when the hot dot hosts one electron, the coupling to the right reservoir is suppressed with respect to coupling to the left reservoir. An electron can therefore be transmitted with higher probability against the electric bias when it is first transmitted from the right reservoir to the work dot while the hot dot is empty. If it remains in the work dot until the hot dot gets populated, its transfer to the left becomes likely. This sequence of events, which is depicted in Fig.~\ref{fig:Engine}(b), leads to a current against the electric bias only if it happens in the correct order. This strongly recalls a thermodynamic cycle, which would involve alternative couplings to baths at different temperatures (the work dot reservoirs vs. the hot dot reservoir) and work exchange (transfer of electron between the work dot and the left or right reservoir electrically biased reservoirs). In order for the steady-state engine to produce power on average, this effective sequence has to happen autonomously during the dynamics of the system and occur more frequently than other sequences, which would result in an opposite transfer of electrons from the left to the right reservoirs. There are many examples of such autonomous engines. Here, the cycle selection is realized by the energy-dependent coupling\footnote{Note that the selection of cycles by energy-dependent couplings can be related to information and feedback in these autonomous machines~\cite{Horowitz2014Jul}. Indeed autonomous Maxwell demons were analyzed theoretically~\cite{Strasberg2013Jan,Sanchez2019Oct} and experimentally~\cite{Koski2015Dec} in these types of setups.}. In some cases a self-oscillation or self-rotation of some degree of freedom of the working fluid was identified as driving the engine cycle~\cite{Alicki2017Mar}.
In the following we show that the engine discussed here does not rely on self-oscillations but on \textit{stochastic} cycles occurring at an increased probability in the steady state.

\section{Ensemble description}\label{sec:average}

We start the description of the dynamics of the energy harvester based on a standard master equation approach, as it was used in Ref.~\cite{Sanchez11} in the original analysis of the setup. Such a description typically captures the dynamics of the \textit{average expectation} value of the double-dot occupations in contact with the three described reservoirs. Based on this, it yields the expectation value of the absorbed heat current averaged over many engine cycles and the resulting produced power, but also second order correlation functions will be evaluated.

\subsection{Master equation and averaged engine performance}

We first recall the master equation treatment of the engine describing the dynamics of the double-dot system in weak contact with the three reservoirs.  
We are interested in the probability of the capacitively coupled double dot being in one of the states $\{ab\}$. We denote this probability as $p_{ab}$. The degrees of freedom of the reservoirs are traced out. The master equation is an equation of motion for these occupation probabilities
\begin{align}\label{eq:master}
  \dot \rho(t) = {\cal W}\rho(t), 
\end{align}
with $\rho(t) = (p_{00},p_{01},p_{10},p_{11})^\mathsf{T}$. 
Note that it is crucial here, that the full dynamics of the double-dot system coupled to reservoirs is entirely described by the diagonal elements of the density matrix of the dots. Coherences (off-diagonal elements) do not enter! The reason for this is the absence of particle tunneling between the two capacitively coupled dots.

The master equation contains a transition rate matrix 
\begin{align}
    \label{eq:ratematrix}
&{\cal W} = \\
&\left(\begin{array}{cccc}
     -W_{0}^+ &  W_{\text{H},0}^- & W_{\text{W},0}^- & 0\\
          W_{\text{H},0}^+ & - W_{\text{W},1}^+ - W_{\text{H},0}^-  & 0 & W_{\text{W},1}^-\\
    W_{\text{W},0}^+ & 0 & -W_{\text{W},0}^- - W_{\text{H},1}^+  & W_{\text{H},1}^-\\
     0 & W_{\text{W},1}^{+} & W_{\text{H},1}^+ & - W_{1}^- 
\end{array}\right)\nonumber
\end{align}
with the following entries found from Fermi's golden rule. Note that in the following we set $\hbar\equiv1$. We write the transition rates for an electron entering ($+$) or leaving ($-$) one of the dots due to tunneling with reservoir $\alpha= \text{L, R, H}$ as 
\bb
\label{eq:trans+}W_{\alpha,\nbar}^+ & =&  \Ganbar f_\alpha(\epsilon_{j_\alpha} + U \nbar),\\
\label{eq:trans-}W_{\alpha,\nbar}^- &  =& \Ganbar (1-f_\alpha (\epsilon_{j_\alpha}+ U \nbar)).
\ee
Here, $f_{\alpha}(\epsilon) = (\exp{\left[\beta_\alpha(\epsilon-\mu_{\alpha})\right]}+1)^{-1}$ is the Fermi-Dirac distribution function of reservoir $\alpha$, with the inverse energy scale set by temperature $\beta_\alpha^{-1}=k_\text{B}T_\alpha$. We furthermore introduce the abbreviation $j_{\alpha} = \text{w}$ for $\alpha = \text{L, R}$, $j_\alpha=\text{h}$ for $\alpha = \text{H}$ to indicate the dot involved in the transition and $\overline{j_\alpha}$ for the dot {\it not} involved in the transition. This means $\epsilon_{j_\text{L}}=\epsilon_{j_\text{R}}=\epsw$, $\epsilon_{j_\text{H}}=\epsh$, and $\nbar \in\{0,1\}$ is the occupation of the hot dot in the case of tunneling to reservoir $\alpha = \text{L, R}$ and the occupation of the work dot in the case of tunneling to reservoir $\alpha = \text{H}$). 
To make notation more concise, we also define  
\begin{align*}
W_{\text{W},\nh}^\pm & = W_{\text{L},\nh}^\pm  + W_{\text{R},\nh}^\pm,\\
W_n^\pm &= W_{\text{W},n}^\pm + W_{\text{H},n}^\pm.
\end{align*}

The master equation, Eq.~(\ref{eq:master}), admits a steady-state solution $\bar \rho = (\bar p_{00},\bar p_{01},\bar p_{10},\bar p_{11})^\text{T}$, verifying ${\cal W}\bar \rho =0$.
The average performance of this engine has already been thoroughly analyzed~\cite{Sanchez11}. We recall here the features useful for our analysis. This engine extracts a steady-state heat current $\bar{J}_\text{H}$ from the hot reservoir 
and converts it into electrical power, $\bar{\cal P}$, namely an electric current flowing against the potential bias. These quantities take the form:

\begin{align}
    \bar{J}_\text{H} &= \sum_{\nw=0,1} (\epsilon_\text{h}+U \nw)\left(W_{\text{H},\nw}^+ \bar{p}_{\nw0}  - W_{\text{H},\nw}^- \bar{p}_{\nw 1} \right),\label{eq:Jheat}\\
\bar{\cal P} &= \Delta\mu\bar{I}_\text{L}= \Delta\mu\sum_{\nh=0,1} \left(   W_{\text{L},\nh}^- \bar{p}_{1\nh} -W_{\text{L},\nh}^+ \bar{p}_{0\nh}\right), \label{eq:power}
\end{align}
with the particle current flowing into the left reservoir $\bar{I}_\text{L}$. The device works as a power generator, when the power $\bar{\cal P}$ is positive, namely when it continuously transfers electrons from the right to the left reservoir against the electric bias $\Delta\mu = \mu_\text{L}-\mu_\text{R}>0$. This regime is found for $\Delta\mu$ lower than a stall bias $\Delta\mu_\text{stop}$ that depends on the various parameters of the problem.

The thermodynamic efficiency of the engine can be computed as 
\bb
\eta = \bar{\cal P}/\bar{J}_\text{H}.
\ee
In the limit ${\cal A}\to 1$ (where the asymmetry is defined in Eq.~\eqref{eq:asym}), the electric and heat currents become proportional to each other, such that~\cite{Sanchez11} the efficiency becomes $\eta \to \eta_\text{ideal} = \frac{\Delta \mu}{U}$. This means that the engine achieves optimal conversion of energy quanta $U$ into electric work quanta $\Delta \mu$ associated with transferring one electron from the right to left reservoir. Within this optimal regime, there is still a trade-off between power and efficiency: when increasing $\Delta\mu$, the power decreases and reaches $0$ for $\Delta\mu = \Delta\mu_\text{stop}$, which corresponds to an efficiency equal to $ \eta_\text{Carnot} = 1-T_\text{W}/T_\text{H}$ the Carnot efficiency. Below, we allow for non-ideal conditions such that $\eta$ can differ from $\eta_\text{ideal}$ and is always smaller than $\eta_\text{Carnot}$.

\subsection{Absence of oscillations in the averaged dynamics}\label{sec:avdyn}

In order to identify whether the dynamics of the double-dot system is characterized by self-oscillations, we rewrite the master equation, Eq.~(\ref{eq:master}), in terms of the occupation numbers of the two separate dots. The average occupations of each dot individually, $\Nw=\langle\hat{n}_\text{w}\rangle$ and $\Nh=\langle\hat{n}_\text{h}\rangle$, are given by $\Nw = p_{10} + p_{11}$ and $\Nh = p_{01}+p_{11}$. Using these expressions, we write the master equation as 
\begin{subequations}
\begin{align}
    \dNw +\Gamma_{\text{W},0}p_{10}+\Gamma_{\text{W},1}p_{11}& =W_{\text{W},0}^+-\left(W_{\text{W},0}^+-W_{\text{W},1}^+\right) \Nh,\label{eq:master_work}\\
    \dNh+\Gamma_\text{H}\Nh&=W_{\text{H},0}^+-\left(W_{\text{H},0}^+-W_{\text{H},1}^+\right)\Nw,\label{eq:master_hot}
\end{align}
\end{subequations}
where we have introduced the abbreviation $\Gamma_{\text{W},\nh}=\Gamma_{\text{L},\nh}+\Gamma_{\text{R},\nh}$. Note that in the case of energy-independent couplings $\Gamma_{\text{W},0}=\Gamma_{\text{W},1}$, we have $\Gamma_{\text{W},0}p_{10}+\Gamma_{\text{W},1}p_{11}=\Gamma_\text{W}\Nw$, leading to two coupled \textit{first-order} differential equations, which do not support any oscillations.

However, the energy dependence of the couplings to the work dot changes this scheme. Then, the equation of motion for the two dot occupations can be cast in two differential equations of \textit{first and second} order. We find
\begin{subequations}
\begin{align}
\ddot{N}_\text{w} + \kappa \dNw + \omega^2 \Nw &= -A \Nh + B,\\
\dNh  + \Gamma_\text{H} \Nh &= -C \Nw  + D,
\end{align}
\end{subequations}
We purposefully write this equation in analogy to the equation of motion of damped, coupled harmonic oscillators driven by external forces,
where the coefficients are defined as
\begin{align*}
\kappa =& \GwN+\GwO+\Gh,\\
\omega^2 =& -\left(W_{\text{W},0}^+-W_{\text{W},1}^+\right)\left(W_{\text{H},0}^+-W_{\text{H},1}^+\right)\\ &+\left(\GwO+\Gh\right)\GwN-(\GwN-\GwO)W_{\text{H},1}^+,\\
A =& \GwO W_{\text{W},0}^+ +\GwN W_{\text{W},1}^+,\\
B =& \left(\GwO+\Gh\right) W_{\text{W},0}^+ - \left(W_{\text{W},0}^+ -W_{\text{W},1}^+ \right) W_{\text{H},0}^+,\\
C =& W_{\text{H},0}^+-W_{\text{H},1}^+,\\
D =& W_{\text{H},0}^+.
\end{align*}
A natural step is now to analyze in what regimes the dynamics of our system has an overdamped or underdamped behavior. These two scenarios correspond to the thermodynamic cycles in the system occurring \textit{stochastically} or \textit{periodically}. 

By taking $y \equiv (\Nw,m,\Nh)$, where $m = \dNw$, we can associate these differential equations with a matrix equation $\dot{y}=Ly+K$, with a $y$-independent part $K$. The eigenvalues of the matrix $L$ correspond to the solutions of the characteristic polynomial (in $\lambda$)
\begin{align}
A C+\Gh  \omega ^2+\lambda  \left(\kappa \Gh +\omega ^2\right)+ \lambda ^2 (\kappa +\Gh )+ \lambda ^3 = 0.
\end{align}
To determine if the solutions of the homogeneous equation $\dot{y}=Ly$ are real or imaginary, we use the discriminant $\Delta$ of this polynomial, which is positive or zero if the solutions are real and negative if the solutions are complex\footnote{For a cubic polynomial $ay^3+by^2+cy+d$, the discriminant is given by 
$\Delta = b^2c^2-4ac^3-4b^3d-27a^2d^2+18abcd$.}. 
We do a numerical search to determine the minimum value of $\Delta$. We therefore fix a functional dependence for the energy dependence of the tunnel couplings to the work dot, namely 
\begin{align}
\Gan = (1-x\delta_{\alpha,\text{R}}\delta_{n,1})\Gamma    .\label{eq:RuleEnergyDep}
\end{align}
We assume this function for the energy dependence of the coupling strengths whenever we present numerical results in the remainder of this paper.  Taking $\epsilon_\text{w}=\epsilon_\text{h}=0$,  and $x=0.9$, we minimize $\Delta$ over $U$, $T_\text{W}$, $T_\text{H}$, and $\Delta \mu$. In addition, we require that $\Delta \mu < U \eta_\text{Carnot}$, which is a necessary condition for the setup to work as an engine (see Eq.~\eqref{eq:NecessaryCond} in Section \ref{sec:cycle_cont}). Using four different numerical methods (Nelder-Mead, differential evolution, simulated annealing, and a random search), we find that the minimum value  $\Delta$ takes on is zero, implying that all the solutions are real. Hence, we conclude that  the average dynamics of this setup do not need to exhibit oscillations despite the fact that power is produced and heat is absorbed from the hot bath. 

\subsection{Correlation functions}\label{sec:corr}

Given the analysis in the above section, we would not expect cycles to occur periodically in the system dynamics. Interestingly, a recent study  suggests that periodically occurring cycles of a steady-state engine, even if hidden in the average dynamics, could be found in the correlation functions of the currents~\cite{Verteletsky20}. The engine analyzed in Ref.~\cite{Verteletsky20} involves two tunnel-coupled superconducting qubits mediating transport of Cooper pairs against an electric bias between two electrodes. The qubits are also coupled to two bosonic baths at different temperatures. Then, even though no oscillations occur in the mean power production, the steady-state correlation functions
were shown to exhibit oscillations as a function of the delay. These oscillations are identified as the strokes of the engine cycle.

Here, we compute analogous quantities in our setup, in order to show that in our case, no oscillations are present in the correlation functions either. Although such two-time correlation functions capture fluctuations around average currents related to single realizations of the experiment, it is possible to compute them directly from the master equation (See e.g. Ref.~\cite{CarmichaelBook}, Chapter 1.5).

\begin{figure}[tb]
\hspace*{-8pt}\includegraphics[width=0.5\textwidth]{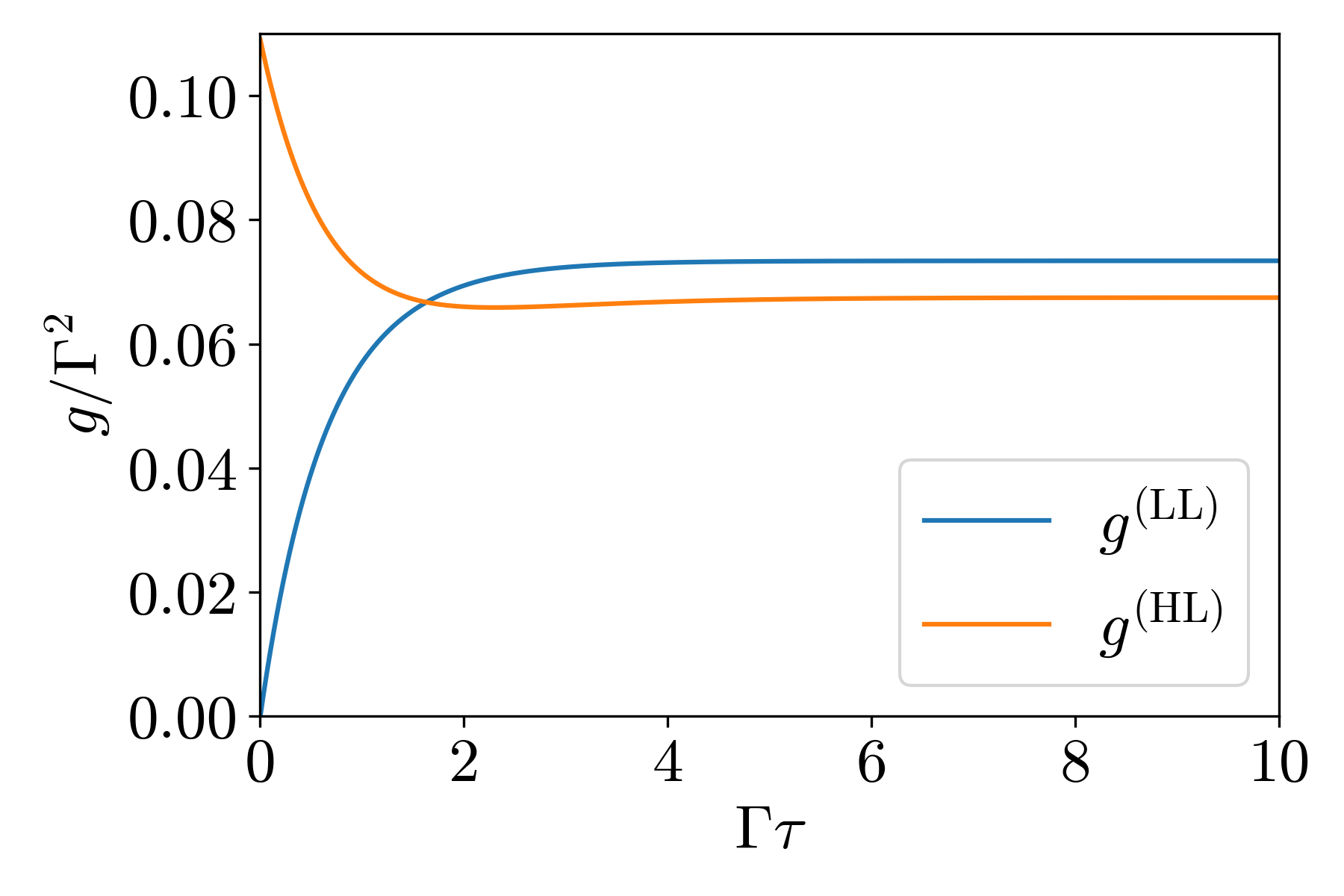}
\caption{Steady-state correlation functions $g^{(\text{LL})}(\tau)$ and $g^{(\text{HL})}(\tau)$ as a function of the delay $\tau$ obtained from numerical solutions of the master equation. We took the energy dependence of the rate given by Eq.~\eqref{eq:RuleEnergyDep} with $x=0.9$, $k_\text{B}T_\text{W}=5\Gamma$, $k_\text{B}T_\text{H} = 15 \Gamma$, $\epsw=\epsh=0$,
$\Delta\mu=\Gamma/4$, and $U=5  \Gamma$.}
\label{fig:ll_corr}
\end{figure} 

We start by defining the steady-state correlation function associated with two consecutive transfers of an electron into the left reservoir $g^\text{(LL)}(\tau)$. In the case of a periodic engine cycle, one can indeed expect that work extraction (here the transfer of electron to the left reservoir), will feature a periodic behavior, leading to (possibly damped) oscillations in $g^\text{(LL)}(\tau)$.
This quantity is given by the steady-state probability of two electron transfers from the work dot into the left reservoir separated by a time delay $\tau$, i.e.:
\bb\label{eq:gLL}
g^\text{(LL)}(\tau) 
= \bar{\pi}(W_\text{L}^-,\tau \vert W_\text{L}^-)  \bar{\pi}(W_\text{L}^-).
\ee
Here, we have introduced the probability $\bar \pi(W_\text{L}^-)$ (per unit of time) that a tunneling event from the work dot to the left reservoir occurs at a time $t$; since the system is in the steady state, $\bar \pi(W_\text{L}^-)$ is independent of $t$. Furthermore, we have introduced the conditional probability $\bar \pi(W_\text{L}^-,\tau \vert W_\text{L}^-)$ (per unit time) that the transition with an electron tunneling out of the work dot into the left reservoir occurs at time $t+\tau$ given that the same type of transition occurred at time $t$. These two probability rates are given by
\begin{align}
    \bar \pi(W_\text{L}^-)& = \sum_{n=0,1} W_{\text{L},n}^- \bar p_{1n},\\
    \bar \pi(W_\text{L}^-,\tau \vert W_\text{L}^-) & = \sum_{n=0,1} W_{\text{L},n}^-  p_{1n}^{\text{L}-}(\tau).
\end{align}
Here, $p_{1n}^{\text{L}-}(\tau)$ is found as elements of the solution $\rho^{\text{L}-}(\tau) = ({p}^{\text{L}-}_{00}(\tau),{p}^{\text{L}-}_{01}(\tau),{p}^{\text{L}-}_{10}(\tau),{p}^{\text{L}-}_{11}(\tau))^\mathsf{T}$ of the rate equation at time $t+\tau$ deriving from the initial condition at time $t$, given by $\rho^{\text{L}-} = \frac{1}{\bar \pi(W_\text{L}^-)} (W_{\text{L},0}^- \bar p_{10},W_{\text{L},1}^- \bar p_{11},0,0)$. This initial condition corresponds to the population of the two-dot system right after the first event, where one electron tunnels out of the work dot into the left reservoir with transition rate $W_{\text{L}}^-$.
The  correlation function $g^\text{(LL)}(\tau)$ is plotted in Fig.~\ref{fig:ll_corr}. 
We compute in an analogous fashion the correlation function $g^\text{(HL)}(\tau)$.  It is associated with the probability of an electron transfer from the hot reservoir into the hot dot, followed by an electron transfer from the work dot into the left reservoir  after a delay time $\tau$, see Appendix~\ref{app:correlation} for details.

Here, the correlation functions show no oscillations, contrary to the two-qubit setup studied in Ref.~\cite{Verteletsky20}. We attribute this difference to the fact that in the system of Ref.~\cite{Verteletsky20} the oscillations in the correlation function reflect the quantum-coherent exchange of excitations which occurs between the two qubits comprising the working substance. Conversely in the present device, the capacitive coupling does not induce any coherences between the dot energy eigenstates. The electronic transfers between reservoirs and the two-dot system studied here are always incoherent, as captured by the rate equation of Eqs.~(\ref{eq:master}) and (\ref{eq:ratematrix}). 
We anticipate that a behavior similar to the setup of Ref.~\cite{Verteletsky20} could possibly be obtained if the mechanism responsible for the energy dependence of the coupling to reservoir R were implemented using a small quantum system able to store an electron and \textit{coherently} exchange it with the upper dot, e.g. another quantum dot. In contrast, the case we currently investigate is obtained when the energy dependence is due to a quantum point contact capacitively coupled to the hot dot~\cite{Sothmann2012May,Yang2019Jul,Mavalankar2013Sep,Maradan2014Jun,Torresani2013Dec,Gasparinetti2012Jun}, i.e. a device which does not host any stable electron level.

Despite the absence of oscillations, Fig.~\ref{fig:ll_corr} shows that correlations can be identified in $g^\text{(LL)}$ and $g^\text{(HL)}$. On one hand, consecutive events of tunneling between left reservoir and work dot are excluded at short time delays~\cite{Emary2012Apr} (a phenomenon similar to the antibunching of photons spontaneously emitted by a driven atom, see e.g. Chapter 2 of Ref.~\cite{CarmichaelBook}). On the other hand, the correlation between tunneling events L$_-$ and H$_+$ is large at short times and decreases at larger delay times $\tau$. 

These results rule out the presence of cycles that occur with a well-defined periodicity in the stochastically generated data. At this point, we can therefore conclude that the underlying engine cycles are followed randomly (in duration and direction). This is in contrast with the paradigm of autonomous mechanical engines characterized by averaged self-oscillations of a degree of freedom playing the role of a piston~\cite{Alicki2017Mar}. 
In the next section, we take a stochastic description to  pinpoint the character of the engine cycles and investigate their role in the engine performance.

\section{Trajectory description}\label{sec:stochastic}

Up to here, we have considered the dynamics of the ensemble-averaged dot occupations. We could thereby exclude that the occurring cycles are \textit{periodic} (namely, there are no self-oscillations), but not directly see the stochastic cycles. In order to get a better understanding of the way in which these stochastic cycles occur, we now approach the problem via a stochastic description.

\subsection{Markov process describing the dynamics}

\begin{figure}[bt]
\begin{center}
\includegraphics[width=0.35\textwidth]{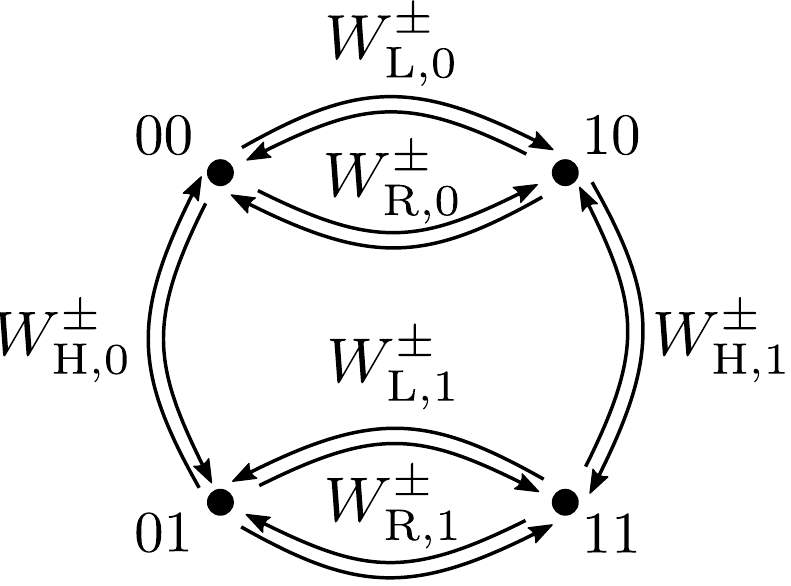}
\end{center}
\caption{Markov chain representing the stochastic dynamics of the double-dot system when electrons exchanged with the reservoirs are counted every time step $\Delta t$. The rates $W_{\alpha,n}^\pm$ set the conditional probabilities for the jumps to occur between times $t$ and $t+\Delta t$ given the system is in the state at the origin of the arrow. In addition to the transitions shown, the system has a non-zero probability to stay in the same state.}
\label{f:Markov}
\end{figure}

We unravel the rate equation for the population under the form of a stochastic process associated with the counting of electrons exchanged with the three reservoirs. This corresponds to a Markov-chain model characterized by the four double-dot states, which are connected by the transition rates involved in the matrix ${\cal W}$, see Eq.~(\ref{eq:ratematrix}). This Markov chain is visualized in Fig.~\ref{f:Markov}. A single realization of the engine operation can be associated with a \textit{trajectory} $\gamma = \{ab_0,(i_1,t_1),(i_2,t_2),...\}$. This is defined by the initial state $ab_0\in\{00,01,10,11\}$ and the jumps $i_\ell \in \{\text{L}_\pm,\text{R}_\pm,\text{H}_\pm\}$. The letters indicate the reservoirs inducing the jump, the subscript determines if the dots receive (+) or provide (-) an electron during the transition, and $t_\ell$ are the times at which the jumps occur.

From the graph describing the Markov chain,  Fig.~\ref{f:Markov}, we see that the trajectories contain different \textit{cycles}. For example,  one can identify six non-trivial cycles starting with the empty state $00$,\footnote{Note that other categorizations of stochastic engine cycles have been put in place, e.g. in Refs.~\cite{Horowitz2014Jul,Lin2020May}.}  which go along with a positive entropy production. 
The different cycles, labeled by ${\cal C}$, are reported in Table~\ref{tab:Cycles}, sorted based on their entropy production \cite{Seifert12,Benenti2017Jun}
\bb\label{eq:entropy}
    \delta\sigma({\cal C}) &=& -\beta_\text{W}(Q_\text{L}({\cal C})+Q_\text{R}({\cal C}))-\beta_\text{H}Q_\text{H}({\cal C})
\nonumber\\
&=&(\beta_\text{W}-\beta_\text{H})Q_\text{H}({\cal C})+\beta_\text{W}\Delta\mu\delta n_\text{L}({\cal C}),\label{eq:dsigC}
\ee
where $Q_{\alpha}({\cal C})$ is the amount of heat provided by reservoir $\alpha =$L,R,H\footnote{Note that in the entropy produced during a cycle $\delta\sigma(\C)$, there is no contribution due to the change of working substance's entropy (i.e. here the double dot system's entropy), as the latter is a state function.} To go to the second line, we have used the first law of thermodynamics, namely $Q_\text{W}({\cal C}) = -Q_\text{H}({\cal C}) - \Delta\mu\delta n_\text{L}({\cal C})$ where $\Delta\mu\delta n_\text{L}({\cal C})$ is the produced electric work and $Q_\text{H}({\cal C})$ the heat received from the hot reservoir during cycle ${\cal C}$. Finally, $\delta n_\text{L}({\cal C}) \in \{\pm 1,0\}$ denotes the number of electrons transferred into reservoir L during cycle ${\cal C}$ (see Table~\ref{tab:Cycles}).

Each of these six cycles can also be traversed in the opposite direction, which we denote with a bar. This leads to opposite heat and work exchanges, and therefore opposite entropy production. The rate $r_{\bar{{\cal C}}}$ of occurrence of the reversed version of a given cycle ${\cal C}$ is linked to the rate of occurrence $r_{\cal C}$ of ${\cal C}$ via (see Chapter 7 of Ref.~\cite{HillBook})
\begin{align}
  \ln\left(\frac{r_{{\cal C}}}{r_{\bar{{\cal C}}}}\right) = \delta\sigma({\cal C}),\label{eq:DBC}
\end{align}
which can be seen as a generalized form of the detailed balance condition, or a special case of the Detailed Fluctuation Theorem \cite{Seifert12}.
Four of the listed cycles contribute to transport between left and right contact; they include two or four jumps, exactly one jump induced by reservoir L and one jump induced by reservoir R.  Only in cycle ${\cal C}_4$, an electron is transferred into reservoir L, as is necessary for the operation of the engine. 

There are three types of modifications of these cycles to be mentioned here. (i) Each of these cycles can be augmented by spurious sub-cycles, for example jumps back and forth the transition $01 \leftrightarrow 11$, which do not contribute to charge or heat transport nor to entropy production. (ii) In Table~\ref{tab:Cycles}, we have furthermore omitted additional sub-cycles L$_+$R$_-$ or L$_-$R$_+$, which can take place after the first and before the last jump of cycles $\mathcal{C}_1$ to $\mathcal{C}_5$ and which would contribute to charge transport. For instance, in the case of cycle ${\cal{C}}_5$, six-state cycles can be formed by adding an L$_+$R$_-$ after the first or third jump.
We denote $\mathcal{C}_1'$ to $\mathcal{C}_5'$ the cycles deriving from cycles $\C_1$ to $\C_5$ when including additional repeated transitions with left and right reservoir without interruptions by transitions with reservoir H. Importantly, cycles $\mathcal{C}_1'$ to $\mathcal{C}_5'$ are very improbable. The reason for this is the large temperature of the hot reservoir, $T_\text{H}>T_\text{W}$, favoring fluctuations. We will demonstrate this suppression in simulations reported in Fig.~\ref{num_cycles}. (iii) Finally, the transitions between states are entirely stochastic and occur with the conditional probabilities given by the rates $W_{\alpha,n}^\pm$. This means in particular that if we divide the trajectory in equally long time segments of duration $\Delta t$, then, after every time step $\Delta t$, there is a finite probability that \textit{no} jump has occurred. Therefore, the duration of the six cycles fluctuates a lot. This will be analyzed in Sec.~\ref{sec:cycle_stat}.

\begin{table}[t]
    \centering
    \begin{tabular}{c||c|c|c|c}
    \hline
       Cycle  & Jumps & $\delta n_\text{L}({\cal C})$ & $Q_\text{H}({\cal C})$ & $\delta\sigma({\cal C})$ \\
   \hline
   \hline
    ${\cal C}_{1}$ & L$_+$H$_+$R$_-$H$_-$ & -1 & $U$ & $(\beta_\text{W}-\beta_\text{H})U+\beta_\text{W}\Delta\mu$\\
   \hline
    ${\cal C}_{2}$ & R$_+$H$_+$R$_-$H$_-$ & 0 & $U$ &$(\beta_\text{W}-\beta_\text{H})U$\\
    \hline
    ${\cal C}_{3}$ & L$_+$H$_+$L$_-$H$_-$ & 0 & $U$ & $(\beta_\text{W}-\beta_\text{H})U$\\
    \hline
    ${\cal C}_{4}$ & R$_+$H$_+$L$_-$H$_-$ & 1 & $U$ & $(\beta_\text{W}-\beta_\text{H})U-\beta_\text{W}\Delta\mu$\\
   \hline
   ${\cal C}_{5}$ & H$_+$L$_+$R$_-$H$_-$ & -1 & 0 & $\beta_\text{W} \Delta \mu$\\
   \hline
   ${\cal C}_{6}$ & L$_+$R$_-$ & -1 & 0 & $\beta_\text{W} \Delta \mu$\\
    \hline
    \end{tabular}
    \caption{Different types of stochastic cycles starting in the empty state $00$ which contribute to the current generation and entropy production. The entropy produced during the cycle reads $\delta \sigma({\cal C}) = (\beta_\text{W}-\beta_\text{H})Q_\text{H}({\cal C})+\Delta\mu \delta n_\text{L}({\cal C})$, where we have introduced the number of electrons coming from the left reservoir $\delta n_\text{L}({\cal C})$ and the heat provided by the hot reservoir $Q_\text{H}({\cal C})$. 
    }
    \label{tab:Cycles}
\end{table}

\subsection{The role of stochastic cycles}

We now show that the steady-state performance of the engine can solely be explained, if the cycles identified above indeed occur during the stochastic double-dot  dynamics. 

Every single trajectory $\gamma$ can be decomposed into a series of cycles starting at 00 and a remaining non-cyclic  piece $\tilde\gamma$. This non-cyclic piece contains the transitions before the system is in the state 00 for the first time and the transitions after the system has been in the state 00 for the last time in a specific trajectory $\gamma$.  In any one cycle ${\cal C}$ in Table~\ref{tab:Cycles}, either 1 or 0 electrons can be exchanged between the work dot and the left reservoir. 
 Furthermore, each cycle ${\cal C}$ occurs in a trajectory $\gamma$ a certain number of times, which we indicate by $N_{\cal C}(\gamma)$.  Finally, $\delta n_\text{L}(\tilde \gamma)$ is the number of electrons transmitted from the work dot to the left reservoir during the remaining non-cyclic  part of the trajectory, $\tilde\gamma$. 
 Due to the possibility of a long sequence of consecutive transitions of the type $\text{R}_+\text{L}_-\text{R}_+\text{L}_-...\text{R}_+\text{L}_-$ while the dot h is occupied, $\vert\delta n_\text{L}(\tilde \gamma)\vert$ can in principle be arbitrarily large. However, the fact that the transitions involving the hot dot have higher probabilities than those involving the work dot and the suppression of the coupling to lead R when $n_\text{h} = 1$ typically restrict $\delta n_\text{L}(\tilde \gamma)$ to $\{\pm 2,\pm 1,0\}$. This is visible from the numerics presnted in Fig.~\ref{num_cycles}.

We now define the stochastic intensity $\cal{I}_\gamma$, provided by the left reservoir along trajectory $\gamma$ of duration $\tau$, and relate it to the generated particle current through the work dot. The stochastic intensity is given by:
\bb
{\cal I}_\gamma = \frac{1}{\tau} \sum_{{\cal C} \in \text{cycles}} N_{\cal C}(\gamma) \delta n_\text{L}({\cal C}) + \frac{\delta n_\text{L}(\tilde \gamma)}{\tau},
\ee
 The sum runs over all different cycles occurring during trajectory $\gamma$. Because $\vert\delta n_\text{L}(\tilde \gamma)\vert \leq 2$ is a reasonable estimate, the last contribution can be neglected for long enough trajectories $\tau \gg \Gamma^{-1}$. 

Using large deviation theory, see Appendix~\ref{app:deviation} and \ref{app:stoch}, it is possible to show that for every sufficiently long trajectory $\gamma$,
\begin{align}\label{eq:intensity_current}
    {\cal I}_\gamma \underset{\Gamma\tau \to \infty}{\sim} \bar{I}_\text{L},
\end{align}
where $\bar{I}_\text{L}$ is the steady state average particle current into the left lead, defined in Eq.~\eqref{eq:power}.
Furthermore, we find for the parameters chosen in Fig.~\ref{fig:ll_corr} 
\begin{align}
    {\cal I}_\gamma \underset{\Gamma\tau \to \infty}{\sim} \bar{I}_\text{L}>0.
\end{align}
Equation~(\ref{eq:intensity_current}) implies that the number of electrons transmitted to the left reservoir is linearly growing with the duration of the trajectory, which means that we must have
\begin{align}
     \bar{I}_\text{L}\simeq \frac{1}{\tau} \sum_{{\cal C} \in \text{cycles}} N_{\cal C}(\gamma) \delta n_\text{L}({\cal C}) =  \sum_{{\cal C} \in \text{cycles}} j_{\cal C}(\gamma) \delta n_\text{L}({\cal C}),\label{eq:ILDF}
\end{align}
with $j_{\cal C}(\gamma) =  N_{\cal C}(\gamma)/\tau$ the probability per unit of time for cycle ${\cal C}$ to occur in the trajectory $\gamma$ (in other words, the stochastic current of cycle ${\cal C}$ in trajectory $\gamma$) \cite{Seifert12}. This quantity is the fluctuating counterpart of the rate of occurrence $r_{\cal C}$ introduced above, i.e. $\moy{j_{\cal C}(\gamma)} = r_{\cal C}$. Eq.~\eqref{eq:ILDF} implies that the long-term fixed intensity of the engine is directly related to the existence of underlying stochastic cycles occurring during the double-dot dynamics.

\subsection{Cycle Statistics}\label{sec:cycle_stat}

We gain more detailed information about the system dynamics by analyzing all cycles occurring in the engine operation during a time interval that is large compared to the average cycle duration. We therefore simulate the system dynamics along long trajectories $\gamma$ and categorize all cycles, which begin and end in the $00$ state, as shown in  Table~\ref{tab:Cycles}.
\begin{figure}[h!]
\centering
\includegraphics[width=0.5\textwidth]{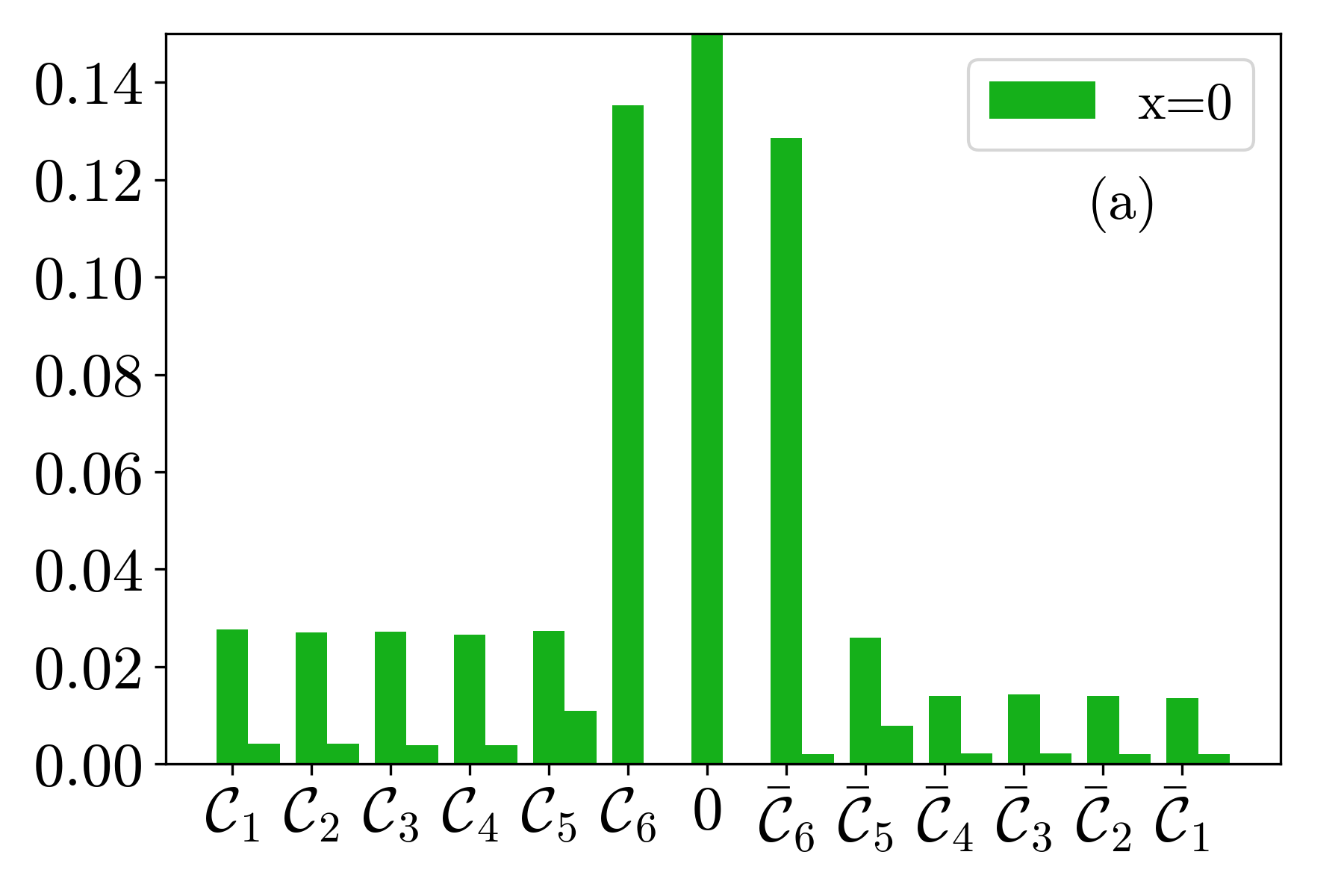}
\includegraphics[width=0.5\textwidth]{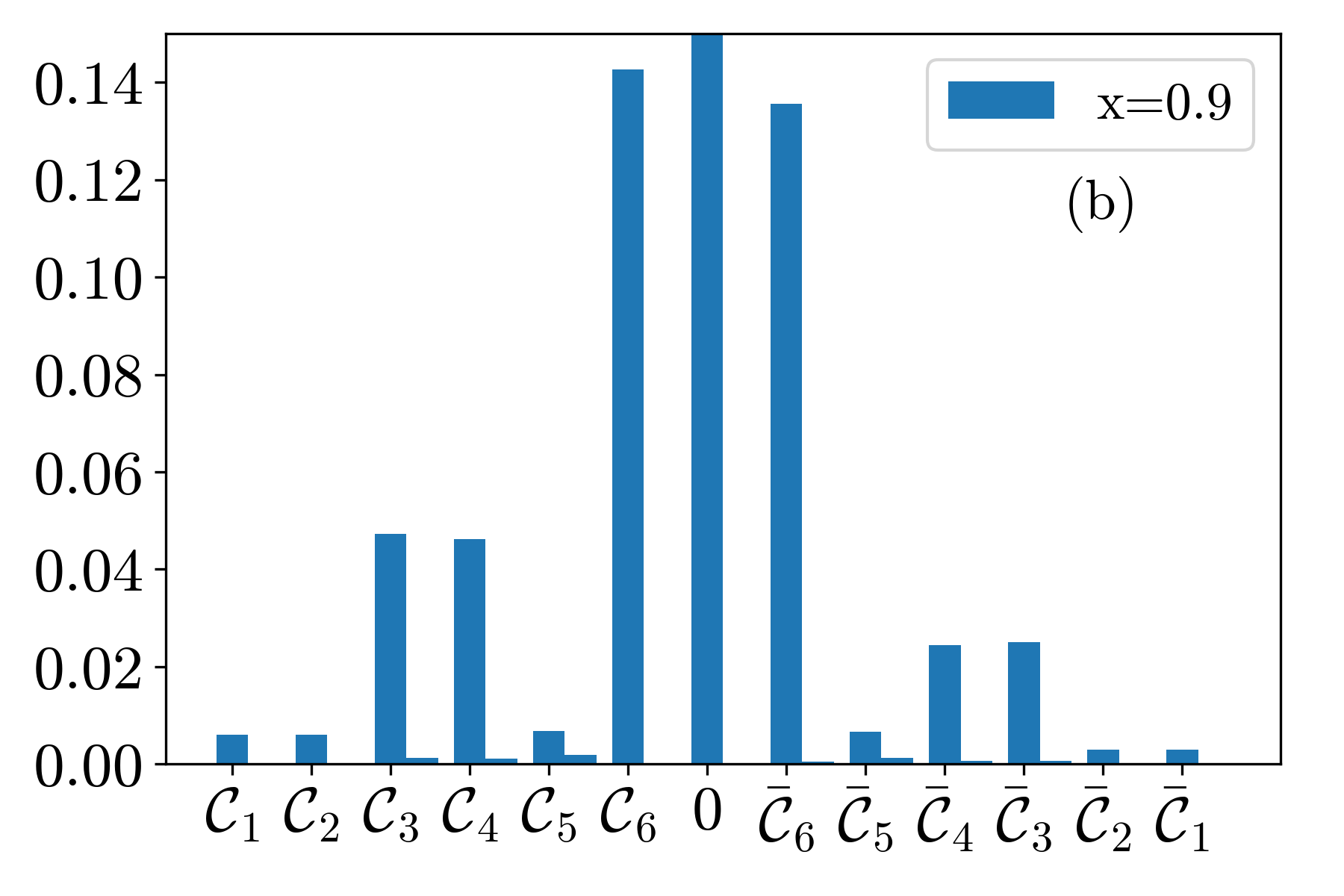}
\hspace*{-13pt}\includegraphics[width=0.515\textwidth]{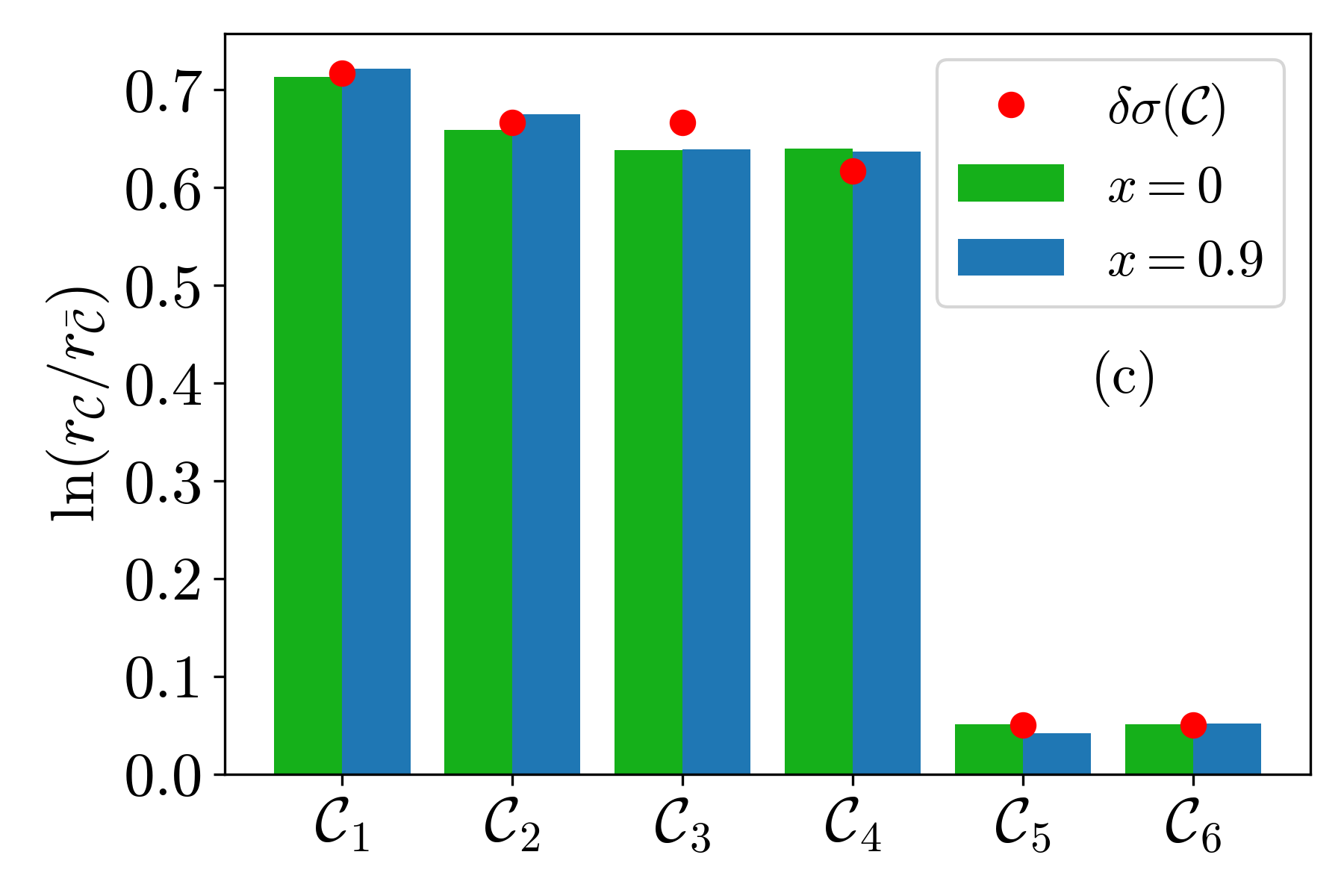}
\caption{(a),(b)~Relative occurrence of possible cycles in a simulation with 10000 trajectories with duration $\tau=20000\Gamma^{-1}$ with (a)~$x=0$ and (b)~$x=0.9$ respectively. (c)~Numerical verification of Eq.~\eqref{eq:DBC}. Red points: $\delta\sigma({\cal C})$ computed from Eq.~\eqref{eq:dsigC}. In contrast, the bars show the quantity $\ln(r_{{\cal C}}/r_{\bar{\cal C}})$ numerically computed from the simulation of 10000 trajectories, when $x = 0$ and $x=0.9$. All plots use parameters $k_\text{B}T_\text{W}=5\Gamma$, $k_\text{B}T_\text{H} = 15 \Gamma$, $\epsw=\epsh=0$, 
$\Delta\mu=\Gamma/4$, and $U=5  \Gamma$.}
\label{num_cycles}
\end{figure} 

We plot the relative occurrence of cycles, $r_\mathcal{C}$, in these trajectories in Fig.~\ref{num_cycles}. The cycle labels correspond to those in Table~\ref{tab:Cycles}, where the cycles are sorted based on their entropy production $\delta\sigma({\cal C})$, see Eqs.~(\ref{eq:entropy}) and (\ref{eq:DBC}). This means that ${\cal C}_1$ has the largest positive entropy production, and ${\cal C}_6$ the lowest. The cycles $\bar{\mathcal{C}}$, followed in reverse direction, shown in the right half of the histogram, Fig.~\ref{num_cycles}, are all less likely than their direct counterpart $\C$. In addition, the unlabeled bars correspond to cycles $\mathcal{C'}$, which are similar to the core cycles $\mathcal{C}$ of Table~\ref{tab:Cycles}, but consist of six transitions or more, typically resulting in $|\delta n_\text{L} ({\cal C})| \geq 1.$ Finally, the central peak (labeled ``0'') corresponds to trajectories in the graph of Fig.~\ref{f:Markov} which come back to the state $00$ without completing any cycle.

To interpret the data presented in Fig.~\ref{num_cycles}, it is most instructive to first look at panel (a), which shows a simulation with energy-\textit{independent} coupling rates. In this case---where no power can be produced---all 4-state cycles $\mathcal{C}_1$ to $\mathcal{C}_5$ occur approximately equally often. Only the two-state cycle $\mathcal{C}_6$ is far more frequent. An equivalent statement applies to the reversed cycles in the right part of the panel. However, only the presence of energy-dependent coupling rates $\Gamma_{\alpha,n}$, allows for finite power production as discussed before.
Hence, the simulation leading to the histogram in panel (b) of Fig.~\ref{num_cycles} involves a transition of rate $\Gamma_{\text{R},1} \ll \Gamma$ captured by a parameter $x=0.9$, see Eq.~(\ref{eq:RuleEnergyDep}).  
Therefore the probability of the cycles ${\cal C}_1$, ${\cal C}_2$ and ${\cal C}_5$, which contain transitions H$_+$R$_+$ and hence involve the coupling rate $\Gamma_{\text{R},1}$, is strongly suppressed. 
The most probable cycles responsible for the engine performance are ${\cal C}_6$ and ${\cal C}_4$. 
Cycle ${\cal C}_4$ is the one depicted in Fig.~\ref{fig:Engine}(b) and it is responsible for the work extraction. Actually, this is the only cycle with four transitions or less, allowing for a current against the bias and a positive entropy production. Conversely, ${\cal C}_6$ corresponds to a work cost that limits the engine performance. Cycle ${\cal C}_3$ does not contribute to the particle current. 
Finally, note that in both histograms Figs.~\ref{num_cycles}(a) and (b), the ratio of the bar height corresponding to a cycle ${\cal C}$ and its time-reverse $\bar{\cal C}$ obey Eq.~\eqref{eq:DBC} as checked numerically in Fig.~\ref{num_cycles}(c). 

\subsection{Properties of the predominant cycle}

\begin{figure}[tb]
\centering
\hspace*{11pt}\includegraphics[width=\columnwidth]{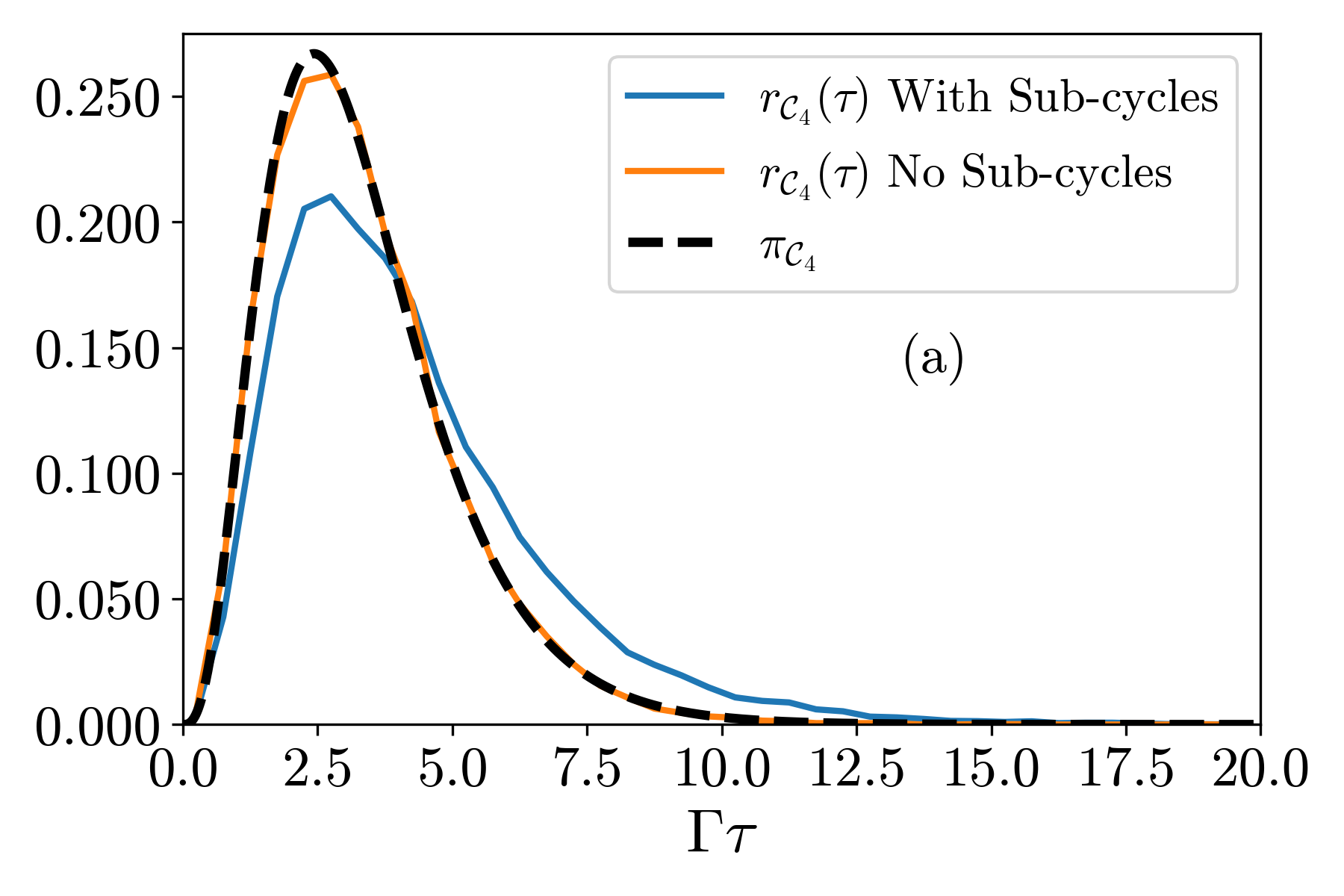}\\
\hspace*{-8pt}\includegraphics[width=1.04\columnwidth]{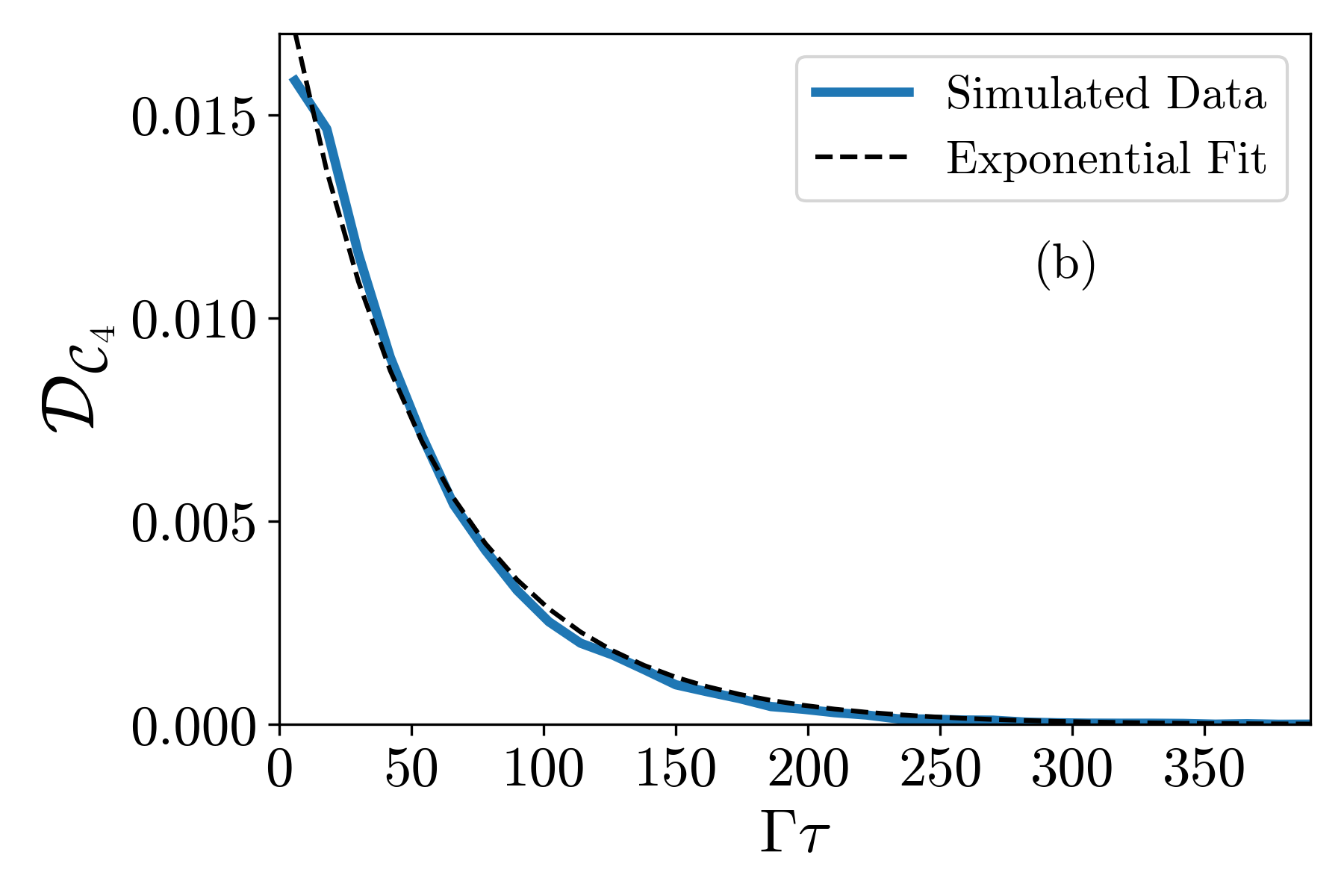}
\caption{(a) Relative occurrence of cycles ${\cal C}_4$ with duration $\tau$, $r_{\mathcal{C}_4}(\tau)$ occurring both with and without back and forth hopping between states. The blue and orange lines represent data collected from the same simulation as in Fig. \ref{num_cycles}, and the black dashed line indicates the distribution found from the time integration of Eq.~(\ref{eq:prob_density}). 
(b) Probability distribution of the time between the beginning of a cycle ${\cal{C}}_{4}$ and the beginning of the next occurrence of the cycle ${\cal{C}}_{4}$ with an exponential fit. Note that the decaying exponential model only applies to times that are long compared to the average length of the cycle, for these parameters $4.4 \Gamma^{-1}$. All panels use the parameters $x=0.9$, $k_\text{B}T_\text{W}=5\Gamma$, $k_\text{B}T_\text{H} = 15 \Gamma$, $\epsw=\epsh=0$, 
$\Delta\mu=\Gamma/4$, and $U=5  \Gamma$.
}
\label{cycle_prob}
\end{figure}

We now focus on the features of cycle ${\cal C}_4$, which is mainly responsible for the power production of our engine. We track the duration of each cycle ${\cal C}_4$ occurring in long trajectories $\gamma$. We plot the relative occurrence of cycles ${\cal C}_4$ with duration $\tau$ with respect to all cycles ${\cal C}_4$ during the steady-state operation of the engine, $r_{\mathcal{C}_4}(\tau)$, in Fig.~\ref{cycle_prob}(a). The  distribution shown by the blue solid line takes into account also all those cycles ${\cal C}_4$ with spurious sub-cycles, since these sub-cycles do not affect the net heat flow and power production. 

Furthermore,  the solid orange line shows the same duration distribution for cycle ${\cal C}_4$, but when it occurs without any additional sub-cycles. We find a peaked distribution, with a maximum occurring at $\tau\simeq3\Gamma^{-1}$.
When taking into account the additional spurious transitions, the probability distribution shifts favoring longer cycles. The peak of the distribution however shifts very little relative to the position of the peak itself. This indicates that while the inclusion of sub-cycles slightly increases the likelihood of iterations of ${\cal C}_4$ with longer duration, the most probable duration is largely determined by the cycles without any additional transitions.

We can confirm the probability distribution of ${\cal C}_4$ by calculating it analytically. To do so, we take the product of the transition probabilities for R$_+$, H$_+$, L$_-$, and H$_-$ at times $t_1$, $t_2$, $t_3$, and $\tau$ in the cycle, where $\tau$ is the return time to the 00 state. The probability of the system remaining in a certain state is described by a decaying exponential, whereas the transitions are accounted for using the definitions in Eqs. (\ref{eq:trans+}) and (\ref{eq:trans-}). 
The probability for the sequence of events R$_+$ at time $t_1$, H$_+$ at time $t_2$, L$_-$ at time $t_3$, and H$_-$ at time $\tau$,  given that the system was in state 00 at time t=0, is then
\begin{align}
    \nonumber &\pi(\text{R}_+,t_1;\text{H}_+,t_2;\text{L}_-,t_3;\text{H}_-,\tau) = e^{-t_1 \left(W_{\text{H},0}^++W_{\text{W},0}^+\right)}W_{\text{R},0}^+\times\\
    &\nonumber    e^{-(t_2-t_1)\left(W_{\text{H},1}^++W_{\text{W},0}^-\right)}W_{\text{H},1}^+ e^{-(t_3-t_2) \left(W_{\text{H},1}^-+W_{\text{W},1}^-\right)} W_{\text{L},1}^- \times \\ 
    &e^{-(\tau-t_3)\left(W_{\text{H},0}^-+W_{\text{W},1}^+\right)}W_{\text{H},0}^- \label{eq:prob_density}
\end{align}
Integrating this over $t_1, t_2,$ and $t_3$ from $0$ to $\tau$, we obtain a probability distribution for the cycle duration $\tau$ of the cycle ${\cal C}_4$ without spurious sub-cycles. We refer to this probability distribution function as $\pi({\cal C}_4,\tau)$. After normalizing by the probability $\int d\tau \pi({\cal C}_4,\tau)$ of occurrence of cycle ${\cal C}_4$ without spurious sub-cycles,  we obtain the probability distribution $\pi_{{\cal C}_4}(\tau)$ of duration $\tau$ for the cycle ${\cal C}_4$.
The agreement with the numerical simulation can be seen in Fig.~\ref{cycle_prob}(a).

Finally, we also track the time difference between the start of two subsequent cycles ${\cal C}_4$. The relative occurrence of the time difference $\tau$ between two subsequent cycles $\mathcal{C}_4$, $\mathcal{D}_{\mathcal{C}_4}(\tau)$, is plotted in Fig.~\ref{cycle_prob}(b). The most probable time difference is close to 0. However, the half life of the curve is rather large, about $ 38 \Gamma^{-1}$ for the parameters given in Fig.~\ref{cycle_prob} (b) when fitted to an exponential curve. These results further support the claim that there is neither a regular, periodic occurrence of the engine cycle ${\cal C}_4$, nor a fixed duration.

\subsection{Contribution of the other cycles}\label{sec:cycle_cont}
 
The analysis of cycle ${\cal C}_4$ alone is enough to get qualitative understanding of the engine operation, but largely overestimates its performances. For instance the stall voltage (value of the electric bias $\Delta\mu$ above which the power vanishes) can be estimated from the necessity of the cycle ${\cal C}_4$ to be mostly run forward, i.e. $r_{\C_4} \geq r_{\Cb_4}$. This leads, using Eq.~\eqref{eq:DBC} to $\delta\sigma(\C_4)> 0$ and therefore to the necessary condition:
\bb
\Delta\mu  < U\eta_\text{Carnot},\label{eq:NecessaryCond}
\ee
with $\eta_\text{Carnot} = 1-\frac{T_\text{W}}{T_\text{H}}$ the Carnot efficiency. For the parameters used in Fig.~\ref{num_cycles}, $k_\text{B}T_\text{W}=5\Gamma$, $k_\text{B}T_\text{H} = 15 \Gamma$ and $U=5  \Gamma$, we find  $U\eta_\text{Carnot} \simeq 3.33\Gamma$. However, numerical simulation show that for the same engine parameter, the power vanishes at a much lower voltage $\Delta\mu_\text{stop}^\text{num} \simeq 0.57\Gamma$ obtained from the full master equation analysis, when solving numerically ${\bar I}_\text{L} \equiv 0$. This means that the constraint set by Eq.~\eqref{eq:NecessaryCond} is not tight.

 This overestimation is due to the contribution of other (detrimental) cycles identified above (in particular $\C_6$, $\C_5$ and $\C_1$ which are detrimental to the engine performance. Below we show that taking into account two cycles ${\cal C}_6$ and ${\cal C}_4$ and their reverse leads to a more accurate analytical prediction of the overall performance of the engine. Using Eq.~\eqref{eq:ILDF}, we take the average over the trajectories $\gamma$ and restrict the sum to only ${\cal C}_4$, ${\cal C}_6$ and their reverse. This leads to:
 \bb
{\bar I}_\text{L} &\simeq& r_{{\cal C}_4}-r_{\bar{\cal C}_4}-r_{{\cal C}_6}+r_{\bar{\cal C}_6}\nonumber\\
&=& r_{{\cal C}_4}(1-e^{-\delta\sigma({\cal C}_4)})-r_{{\cal C}_6}(1-e^{-\delta\sigma({\cal C}_6)}),
\ee
where we have used $\moy{j_\C(\gamma)}=r_\C$ and Eq.~\eqref{eq:dsigC}. We see that work is extracted on average as long as
\bb
\frac{r_{{\cal C}_4}}{r_{{\cal C}_6}} \geq \frac{1-e^{-\delta\sigma({\cal C}_6)}}{1-e^{-\delta\sigma({\cal C}_4)}} = \frac{1-e^{-\beta_\text{W}\Delta\mu}}{1-e^{\beta_\text{W}\Delta\mu-(\beta_\text{W}-\beta_\text{H}) U}}.\label{ineq}
\ee

The ratio $\frac{r_{\C_4}}{r_{\C_6}}$ can also be evaluated using the tools of graph theory \cite{HillBook}. We find (See Appendix \ref{app:cyclecurr}):
\bb\label{ratio}
\frac{r_{\C_4}}{r_{\C_6}} = \frac{W_{\text{R}0}^+ W_{\text{H}1}^+ W_{\text{L}1}^- W_{\text{H}0}^-}{W_{\text{L}0}^+ W_{\text{R}0}^-(W_{\text{H}0}^- W_{\text{H}1}^- + W_{\text{W}1}^- W_{\text{H}0}^- + W_{\text{W}0}^+ W_{\text{H}1}^-)}.\nonumber\\
\ee

Using Eqs.~\eqref{ineq} and \eqref{ratio}, we find a condition for the stall voltage (the bias appears both in the right-hand sides of Eqs.~\eqref{ineq} and and \eqref{ratio}). For the parameters used in Fig.~\ref{num_cycles}, namely $x=0.9$, $k_\text{B}T_\text{W}=5\Gamma$, $k_\text{B}T_\text{H} = 15 \Gamma$, $\epsw=\epsh=0$, and $U=5  \Gamma$, we find  $\Delta\mu \leq \Delta\mu_\text{stop} \simeq 0.69\Gamma$, which is much closer to the numerical value $\Delta\mu_\text{stop}^\text{mum} \simeq 0.57\Gamma$ than the previous estimate derived from Eq.~(\ref{eq:NecessaryCond}). The fact that the result from Eq.~(\ref{ineq}) is still a slight overestimation, shows the importance of other detrimental cycles --- like ${\cal C}_1$, $\C_5$ --- for these parameters. Their contribution can be taken into account explicitly with the same methods (See Appendix \ref{app:cyclecurr}).

\subsection{Work extraction mechanism}\label{sec:noncons}

We have shown that in this engine, no degree of freedom plays the role of a work-exchanging piston undergoing well defined oscillations. It is therefore not possible to define an effective time-dependent Hamiltonian for the working dot whose time derivative would yield the power produced by the engine. It is interesting to characterize further the work extraction mechanism occurring here. Each electron transferred from the work dot to the left lead corresponds to an amount of electrical work extracted $\Delta\mu$ \cite{Benenti2017Jun}. In sharp contrast with the work performed on a system by driving it in a controlled way, which can be quantified from the time-derivative of a time-dependent Hamiltonian \cite{Alicki79}, here an amount of work is exchanged during each stochastic transition triggered by the lead L. When an electron is transferred from the dot w to the lead L, the energy variation of dot w is $\tilde\epsilon_\text{w} +n_\text{h}U$. However, only the amount $\tilde \epsilon_\text{w}+n_\text{h}U-\mu_\text{L} = \epsilon_\text{w}+n_\text{h}U - \Delta \mu$ corresponds to heat. Only this splitting ensures the positivity of the average entropy production \cite{Benenti2017Jun}. Namely, this prescription leads us to define the heat flows $Q_{\alpha}(\C)$, and then the entropy production per cycle $\delta \sigma({\cal C})$ (summarized in Table~\ref{tab:Cycles}), ensuring positive entropy production in average:
\bb
\dot\sigma = \sum_{{\cal C}\in\text{cycles}}r_{\cal C}\delta \sigma({\cal C}) \geq 0.
\label{entropy-increase}
\ee

This behavior indicates that the leads play a double role here, the one of thermal reservoirs at thermal equilibrium, and the one of a battery able to store or provide work. The latter role requires a constant generalized force $F_\alpha$ to be applied on the reservoir, causing a deviation from the canonical distribution and therefore a modified local detailed balance \cite{Horowitz16}  $W_{\alpha,n_\text{h}}^-/W_{\alpha,n_\text{h}}^+ = e^{\beta_\alpha(\epsilon_\text{w}+n_\text{h}U)}e^{-\beta_\alpha F_\alpha}$, for $\alpha = L,R$. Here the constant force comes from the electric bias and we can identify $F_\alpha = \Delta\mu \delta_{\alpha,\text{L}}$. A similar work exchange mechanism can occur in different contexts, e.g. a spatially translated thermal reservoir \cite{Horowitz16}, a squeezed thermal reservoir \cite{Manzano18}, or the electromagnetic reservoir of an atom in presence of strong quasi-resonant drive \cite{Elouard20}.

We note that in addition to the second law of thermodynamics being satisfied (\ref{entropy-increase}), the cycle-resolved entropy production also satisfies the integral fluctuation theorem,
\bb
\langle e^{-\delta \sigma} \rangle = 1,
\ee
where the average is taken over cycles \cite{crooks1999entropy}.  This can be seen explicitly as
\begin{eqnarray}
\langle e^{-\delta \sigma} \rangle &=& \frac{\sum_{\cal C} r_{\cal C} \exp(-\delta \sigma({\cal C}))}{{\sum_{\cal C} r_{\cal C}}},  \\
&=& \frac{\sum_{\cal C} r_{\bar{{\cal C}}} }{\sum_{\cal C} r_{\cal C} } = 1.
\end{eqnarray}
Here, we have expressed the average over cycle probabilities as the sum over the relative occurrences, normalized by all relative occurrences.  The second lines comes from the definition of the entropy production, Eq.~(\ref{eq:DBC}), while the final equality comes from the fact that every cycle has a time-reversed cycle in the sum.

\section{Semi-Stochastic Description}\label{sec:semi}

Above, we have shown that the \textit{work extraction} mechanism is not related to a time-dependence in the Hamiltonian of dot w. However, remarkably, the \textit{heat injection} mechanism can be expressed as a \emph{stochastic} time-dependent term in the effective Hamiltonian of dot w. In this last section, we develop a semi-stochastic description of the dynamics of dot w which highlights that the dot h and the hot reservoir can be seen as a \emph{stochastic} piston exchanging heat. For the sake of simplicity, we focus on a limit, where there is no backaction of the dynamics of the work dot on the dynamics of the hot dot. This happens when $T_\text{H} \gg U$, such that we have that $W_{\text{H},0}^\pm \approx W_{\text{H},1}^{\pm}$. We then label these transitions as $W_\text{H}^\pm$. In this backaction-free regime, the master equation for the hot dot, see Eq.~(\ref{eq:master_hot}), becomes
\begin{align}
    \dot{N}_\text{h}  + \Gamma N_\text{h}&= W_\text{H}^+.
\end{align}
\begin{figure}[bt]
\centering
\includegraphics[width=\columnwidth]{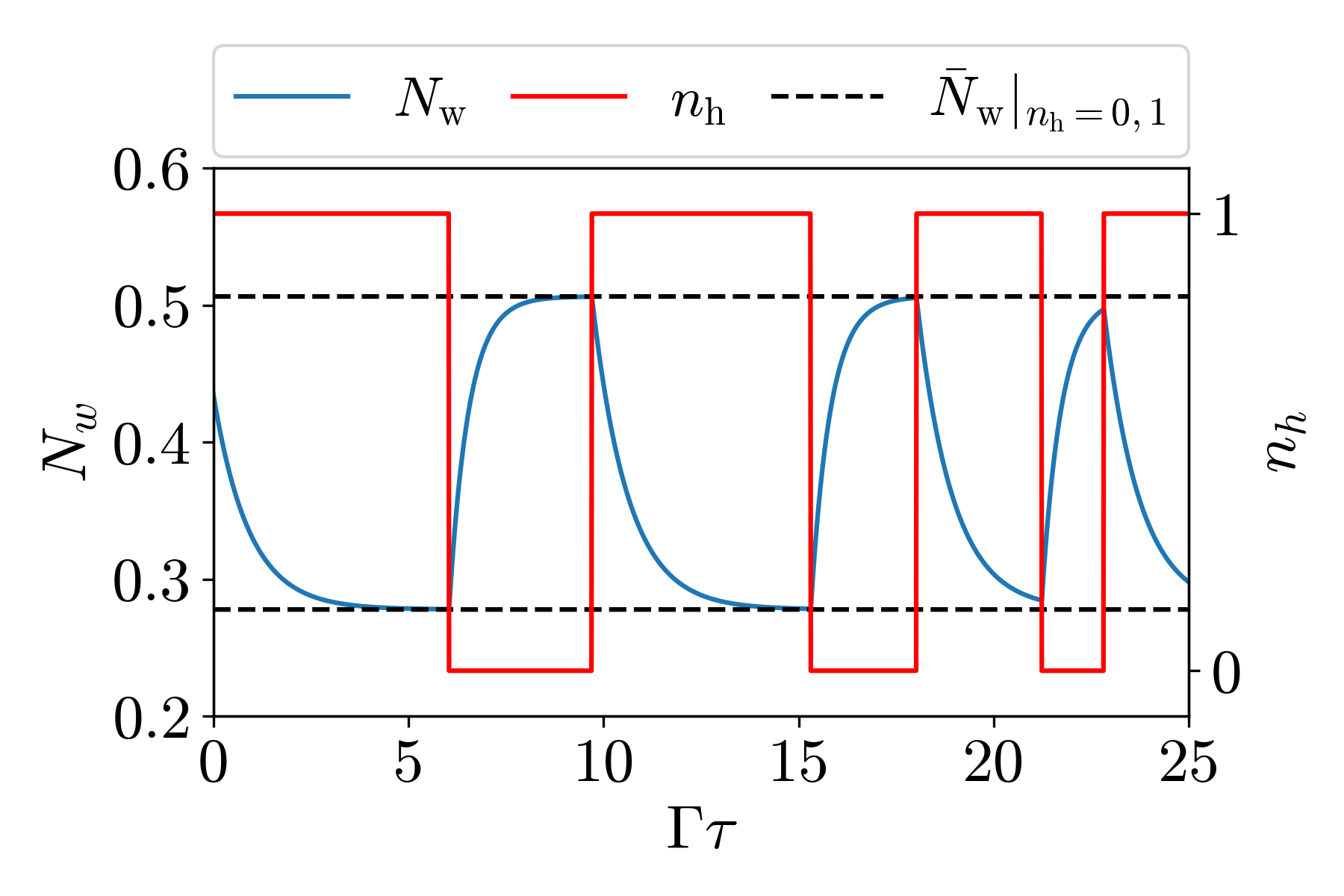}
\caption{Occupation $N_\text{w}$ of the work dot in the semi-stochastic model together with the stochastic changes in $n_\text{h}$. Following the master equation, Eq.~(\ref{eq:master_semi}), the occupation $N_\text{w}$ increases when $n_\text{h}$ is not occupied and decreases when it is occupied.}
\label{f:semi-stoch}
\end{figure}
As argued before, it is independent of the occupation of the work dot, $N_\text{w}$. We can therefore treat the hot-dot occupation as a random variable, $N_\text{h}\rightarrow n_\text{h}$ , that adjusts the occupation energy of the upper dot by
\begin{align}
    \epsilon_\text{w}(t) = \epsilon_\text{w}+U n_\text{h}(t),
\end{align}
leading to the time-dependent effective Hamiltonian of the work dot,
\bb
H_\text{w}(t) = (\epsw + U \nh(t))\hat{n}_\text{w}.
\ee
Here $\nh(t)$ switches stochastically between $0$ and $1$, such that this scenario has a striking resemblance to work extraction from telegraph noise~\cite{Entin-Wohlman17}.
The dynamics of the average work-dot occupation $N_\text{w}$ is instead described by a master equation depending on the stochastic variable $n_\text{h}(t)$. It is obtained from Eq.~(\ref{eq:master_work}) by noticing that in the absence of backaction, we can write $p_{10}=N_\text{w}(1-n_\text{h})$ and $p_{11}=N_\text{w}n_\text{h}$. We find
\begin{align}
    \nonumber \dot{N}_\text{w} + \Gamma_{\text{w},0} N_\text{w} - (\Gamma_{\text{w},0}-\Gamma_{\text{w},1}) N_\text{w} n_\text{h}  \\
    = -\left(W_{\text{w},0}^+ - W_{\text{w},1}^+\right) n_\text{h}  + W_{\text{w},0}^+.\label{eq:master_semi}
\end{align}
Note that, in the limit $k_\text{B}T_\text{H}\gg U$, the solution of Eq.~\eqref{eq:master_work} can be retrieved by averaging the solution of  Eq.~\eqref{eq:master_semi} over many realizations of the hot dot stochastic dynamics.

In Fig.~\ref{f:semi-stoch}, we plot a single realization of the dynamics of the occupation probability $N_\text{w}$ of the work dot together with the stochastic jumps in $n_\text{h}$. In this picture, the engine follows cycles of fixed direction (just as in Fig.~\ref{fig:Engine}(a). The four strokes can be identified as (i) $\Nw$ decreases while $\nh = 1$, (ii) $\nh$ jumps from 1 to 0, (iii) $\Nw$ increases while $\nh = 0$, (iv) $\nh$ jumps from 0 to 1. However, this cycle occurs with a \emph{stochastic amplitude}, defined as the difference between the maximum and minimum values of $\Nw(t)$ over one cycle. 
Indeed, between two transitions in the hot dot, the work dot population evolves according to the master equation Eq.~\eqref{eq:master_semi}. As the delay between consecutive transitions of the hot dot is random, a different population for the work dot is reached. For long enough delays, the steady state values
\bb
\bar{N}_\text{w}\vert_{n_\text{h}=0} = \frac{W_{\text{W},0}^+}{\Gamma_{\text{W},0}},\nonumber\\
\bar{N}_\text{w}\vert_{n_\text{h}=1} = \frac{W_{\text{W},1}^+}{\Gamma_{\text{W},1}}
\ee 
are reached (see Fig.~\ref{f:semi-stoch}). Remembering that exactly one transition $\text{H}_+$ and one transition $\text{H}_-$ occur during the cycle, one can easily compute the heat transferred into the work dot as:
\begin{align}
    q_\text{in} = U\Delta N_\text{w},
\end{align}
where $\Delta n_\text{w}=N_\text{w}(t_1)-N_\text{w}(t_1+t_2)$ is the stochastic amplitude of the cycle, i.e. the difference between the occupation of the work dot at the time $t_1$, where the step $n_\text{h}=0\to1$ takes place, and the occupation of the work dot at time $t_2$, where the transition $n_\text{h}=1\to0$ takes place. The maximum value which this difference can take is $\bar{N}_\text{w}\vert_{n_\text{h}=1}-\bar{N}_\text{w}\vert_{n_\text{h}=0}$.
 
\begin{figure}[tb]
\centering
\includegraphics[width=\columnwidth]{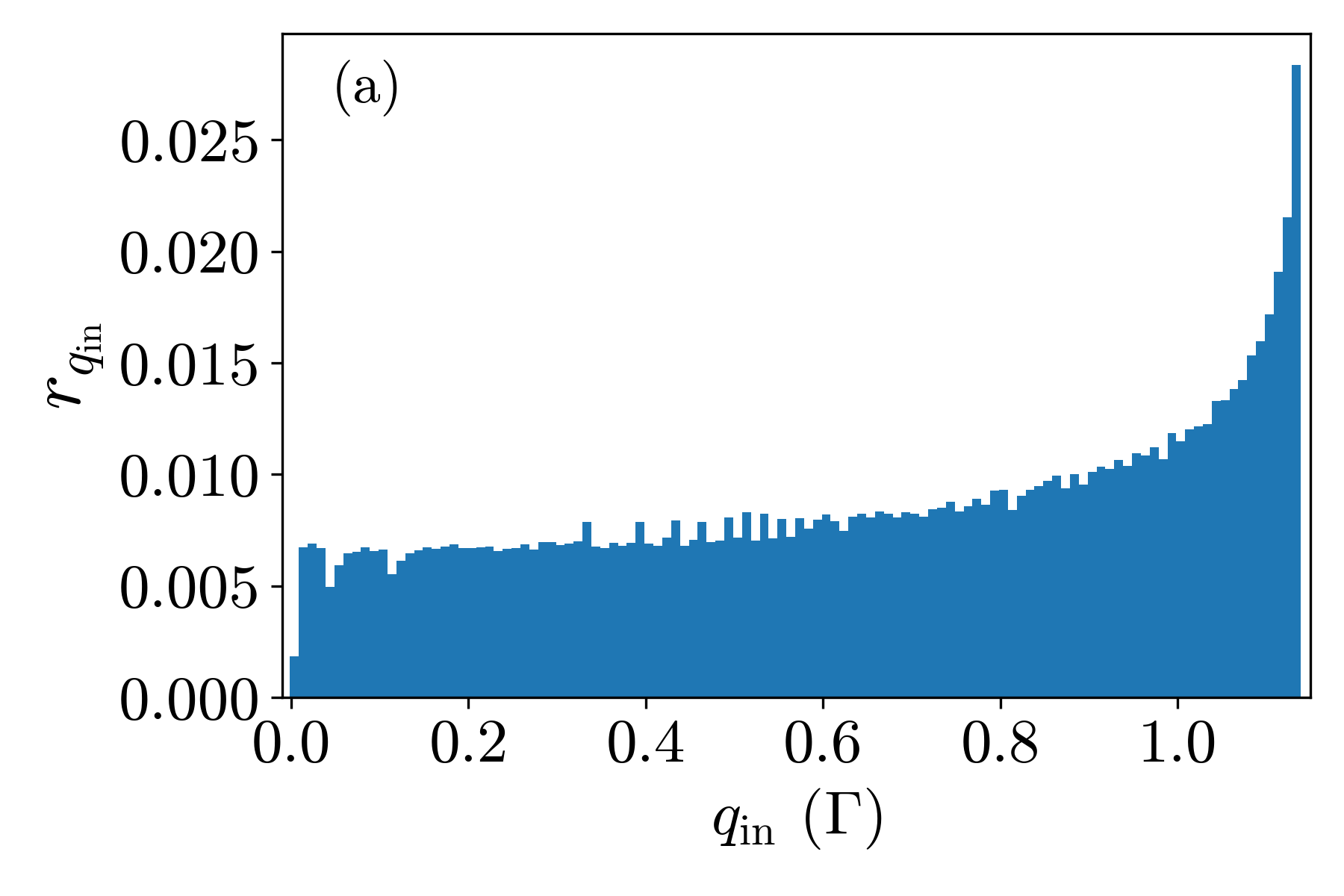}
\includegraphics[width=\columnwidth]{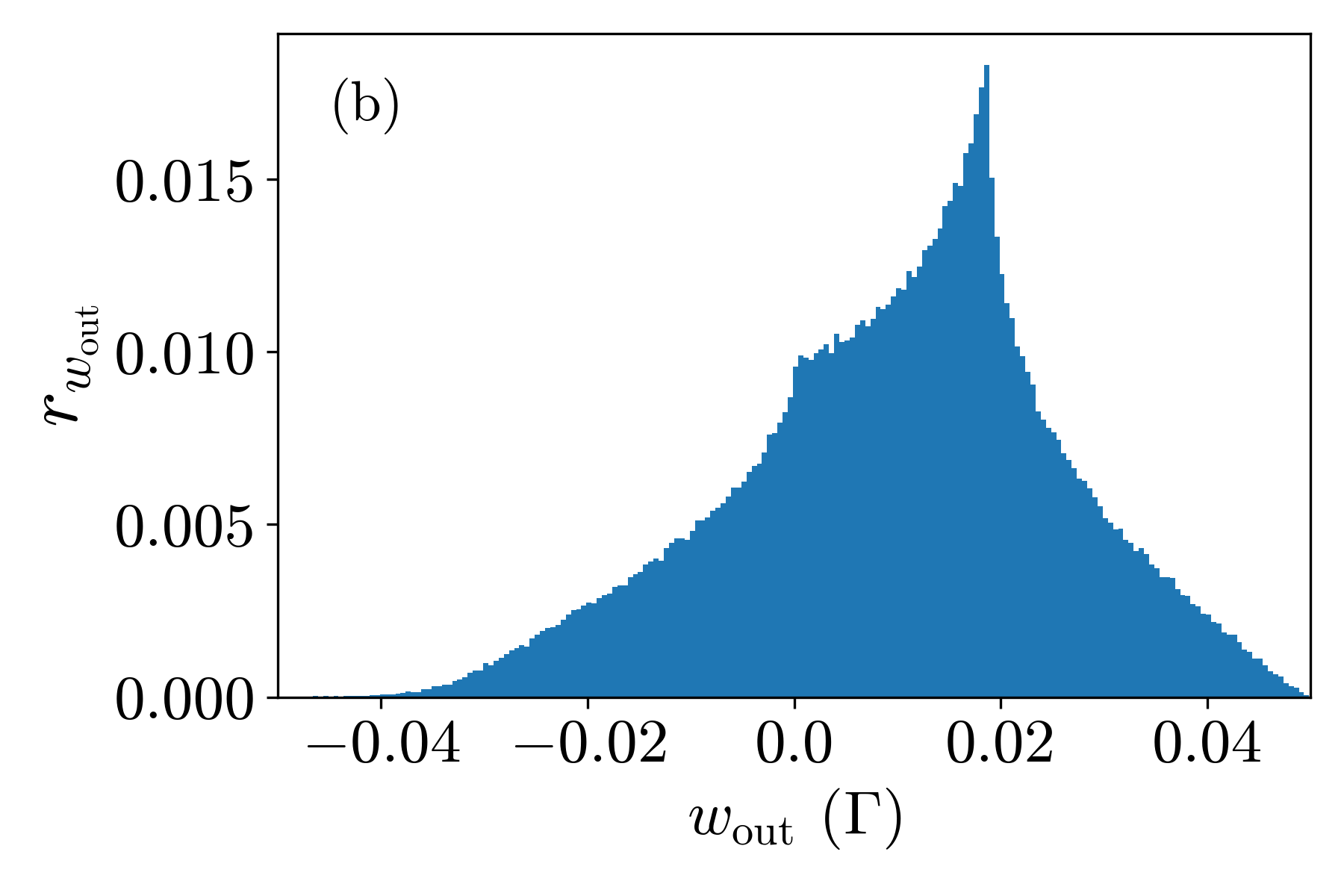}
\caption{Histogram for the occurrence of (a) $q_\text{in}$ transferred and (b) work done in a single cycle, as determined from a semi-stochastic simulation over 1000 trajectories each lasting $2000 \Gamma^{-1}$.  We use the parameters $x=0.9$, $k_\text{B}T_\text{W}=5\Gamma$, $k_\text{B}T_\text{H} = 100 \Gamma$, $\epsw=\epsh=0$,
$\Delta\mu=\Gamma/4$, and $U=5  \Gamma$. The value $\text{max}[q_\text{in}]=1.14\Gamma$ found in this simulation coincides with the analytical result Eq.~(\ref{eq:qmax}).}
\label{f:heat_transfer}
\end{figure}
With this we determine the heat transfer in each cycle occurring during a long stochastic trajectory and plot its relative occurrence, $r_{q_\text{in}}$, as a histogram in Fig.~\ref{f:heat_transfer}. The peak on the right-hand side of the graph corresponds to the maximum heat transfer that occurs over the course of a cycle. 
This maximum value $\text{max}[q_\text{in}]$ can be found by calculating the difference between the steady-state value of $N_\text{w}$ when the hot dot is occupied or empty. We find that 
\begin{align}
    \label{eq:qmax} \text{max}[q_\text{in}] = U \left(\frac{W_{\text{W},0}^+}{\Gamma_{\text{W},0}} - \frac{W_{\text{W},1}^+}{\Gamma_{\text{W},1}} \right).
\end{align}
The fact that this maximum value is also the most probable heat transferred indicates that for the chosen parameters, the work dot can typically come close to its steady-state value before a transition can occur in the hot dot. \par To calculate the work done in this model, we integrate the equation for power, Eq.~(\ref{eq:power}), over the length of a cycle. The cycle begins when the hot dot is empty at time $t_0$. The cycle continues when the hot dot is occupied at time $t_1$ and ends when the hot dot empties at $t_2$. We also assume that all of the energy dependence is contained in the right lead so that $\Gamma_{\text{L},0} = \Gamma_{\text{L},1} = \Gamma_\text{L}$. In this case, the extracted work is
\bb
\nonumber w_\text{out} = \Delta \mu \left(\int \limits_{t_0}^{t_2} \Gamma_\text{L} N_\text{w}(t) \, dt - W_{\text{L},0}^+ \Delta t_0 - W_{\text{L},1}^+ \Delta t_1\right),
\ee
where $\Delta t_0 = t_1 - t_0$ (resp. $\Delta t_1 = t_2 - t_1$) is the time of the cycle during which $n_\text{h} = 0$ (resp. $n_\text{h} = 1$). To eliminate the integral, we solve Eq.~(\ref{eq:master_semi}) for $N_\text{w}$ and substitute the result into the above equation. With this, we find 
\bb
\nonumber w_\text{out} = \Delta \mu \sum_{j = 0,1} &\frac{\displaystyle \Gamma_\text{L}}{\displaystyle \Gamma_{\text{W},j}}\left(N_\text{w}(t_{j}) - N_\text{w}(t_{j+1}) \right) \\
&+\Delta t_{j} \left(\frac{\displaystyle \Gamma_\text{L}}{\displaystyle \Gamma_{\text{W},j}} W_{\text{W},j}^+ - W_{\text{L},j}^+ \right)
\ee

Note that here the label $j$ stands for the hot-dot occupation when it enters as a subscript of the rates. In the time-interval $[t_j,t_{j+1}]$ the hot dot is empty if $j=0$ and occupied if $j=1$.
We plot the relative occurrence of work done, $r_{w_\text{out}}$ in the same manner as the heat transfer. 
\par This effective time-dependent description, together with Fig.~\ref{f:semi-stoch}, are therefore another way to highlight the cyclic but stochastic nature of the process underlying the operation of this energy harvester.

As a last comment, we highlight that the heat current provided to dot w by the hot source can also be computed as the expectation value of $\dot{H}_\text{w}(t)$, where $H_\text{w}(t) = (\epsilon + n_\text{h}(t)U)\hat{n}_\text{w}$. While this term looks like the power provided to a quantum system by a drive~\cite{Alicki79}, here the corresponding energy change behaves as heat due to the stochastic nature of the evolution of $n_\text{h}(t)$. This example highlights the subtlety of the thermodynamic interpretation of energy exchanges in autonomous systems.

\section{Conclusion}\label{sec:conclusions}

We have analyzed a thermoelectric, steady-state three-terminal energy harvester with respect to possible cycles occurring in its dynamics. This energy harvester, based on two Coulomb-coupled quantum dots, extracts heat from a hot reservoir to induce an electric current between two biased colder reservoirs. 
Our detailed analysis shows that stochastic cycles in state space of the double quantum dot are underlying the working principle of this engine. At the same time we have shown the cycles not to be regular in length and direction. This means in particular that these cycles are not periodic.

More specifically, we have analyzed the dynamics of the dot occupation numbers, using a master equation approach. These dynamics, described by first- and second-order differential equations show an overdamped behavior. This goes along with the fact that also correlation functions between different tunneling contributions do not show oscillations. Such behavior is in contrast to steady-state engines relying on quantum-coherent superpositions, where such oscillations were shown to occur in correlation functions~\cite{Verteletsky20}. 

From a different perspective, we described the trajectory dynamics of the system using a stochastic approach. Indeed, the stochastic dynamics features a distribution of cycles occurring randomly in duration and direction, which we showed to be essential to explain the engine performance. We characterized the cycle responsible for the main contribution to the engine performance, and showed it occurs at random times and has a fluctuating duration. We related the rate of occurrence of the various cycles to the entropy produced during each of them.

Finally, we have argued that such an autonomous engine without self-oscillations relies on a non-unitary, dissipative work extraction mechanism, driven by the coupling to reservoirs. Such work exchange is only possible if the reservoirs are subject to a constant generalized force (here the electric bias).
 What is more, we have shown that the heat transfer from the hot source, via the dot coupled to it, can be represented as the action of a stochastic piston. This stochastic piston acts on the dot responsible for work extraction. In our theoretical description it enters as a randomly varying Hamiltonian of this dot. 
 Despite the fact that a time-varying Hamiltonian is often associated to work exchange, here the stochasticity of the time-dependence imposes to classify this energy exchange as heat.
 Our analysis illustrates that the splitting between heat and work can be more subtle than the difference between unitary (Hamiltonian) and non-unitary (dissipative) dynamics, and open new perspectives about the connection between cyclic and steady-state engines.

\acknowledgments
We acknowledge stimulating discussion with the participants of the KITP program ``Thermodynamics of Quantum Systems", where this work was started. We thank R. S\'anchez for helpful comments on the manuscript.  Work by CE and ANJ was supported by the US Department of Energy (DOE), Office of Science, Basic Energy Sciences (BES), under Grant No. DE-SC0017890. JS acknowledges financial support by the Knut and Alice Wallenberg foundation and the Vetenskapsr\r{a}det (Swedish  VR). This work was supported in part by the National Science Foundation under Grant No. NSF PHY-1748958. 

\appendix

\section{Correlation function}\label{app:correlation}

In addition to the correlation function $g^{(\text{LL})}(\tau)$, given in Eq.~(\ref{eq:gLL}), we also show the correlation function $g^\text{(HL)}(\tau)$ in Fig.~\ref{fig:ll_corr}. In this appendix, we show the definition of $g^\text{(HL)}(\tau)$. Namely, it is defined as:
\begin{align}
g^\text{(HL)}(\tau) &= \bar p(W_\text{L}^-,\tau\vert W_\text{H}^+) \bar p(W_\text{H}^+).
\end{align}
Here, we have introduced two different probabilities per unit time, defined by 
\begin{align}
\bar \pi(W_\text{H}^+) & = \sum_{n=0,1} W_{\text{H},n}^+ \bar p_{n0}\\
\bar \pi(W_\text{L}^-,\tau \vert W_\text{H}^+) & = \sum_{n=0,1} W_{\text{L},n}^- \bar p_{1n}^\text{H+}(t)
\end{align}
The second expression, namely for the conditional probability that an event L$_-$ occurs at a time $\tau$ after an event H$_+$ has occurred,  contains the probabilities $\bar p_{1n}^\text{H+}(t)$. They are found as elements of the solution of the rate equation $\rho^\text{H+}(\tau) = ({p}^\text{H+}_{00}(\tau),{p}^\text{H+}_{01}(\tau),{p}^\text{H+}_{10}(\tau),{p}^\text{H+}_{11}(\tau))^\mathsf{T}$, deriving from the initial condition $\rho^\text{H+}(0) = (0,W_{\text{H},0}^+ \bar p_{00}(t),0,W_{\text{H},1}^- \bar p_{10}(t))/\bar \pi(W_\text{H}^+)$. This initial condition corresponds to the populations of the two-dot system right after the event $W_{\text{H}}^+$.\\

\section{Large deviation theory and counting statistics}\label{app:deviation}

In order to keep track of the fluctuations in the particle current $I_\text{L}$ into the left reservoir and the heat flow $J_\text{H}$ from the hot reservoir, one can use the counting statistics method. We are interested in characterizing the probability density $P(t,I,J)$ of having the time-averaged particle current belonging to $[I,I+dI]$ and time-averaged heat current in $[J,J+dJ]$ after operating the engine during $t$. Large deviation theory predicts that for sufficiently long times $t \gg \Gamma^{-1}$ this probability has the form 
\bb\label{eq:PIJ}
P(t,I,J) = C(t,I,J) e^{R(I,J)t}
\ee
where $\displaystyle\lim_{t\to\infty} \tfrac{1}{t}\ln C(t,I,J) = 0$. The function $R(I,J)$ is called the large deviation function and is a valuable tool to understand the long time behavior of the system beyond the average analysis. The large deviation function can be deduced from the master equation using a counting statistics approach \cite{Bagrets02,Sukhorukov07,Esposito09}. 

As a first step, we introduce the generating function $G(t,\lambda,\xi) = \int_{-\infty}^\infty dI  \int_{-\infty}^\infty dJ e^{-(\lambda I + \xi J)t}P(t,I,J)$ that was already determined earlier for the current setup~\cite{Sanchez12,Sanchez13}. It can be computed as $G(t,\lambda,\xi) = \sum_{ab} p_{ab}(t,\lambda,\xi)$ where the populations $p_{ab}(t,\lambda,\xi)$ are the solutions of the modified master equation $\dot \rho(t,\lambda,\xi)= {\cal W}(\lambda,\xi)\rho(t,\lambda,\xi)$, with $\rho(t,\lambda,\xi) = (p_{00}(t,\lambda,\xi),p_{01}(t,\lambda,\xi),p_{10}(t,\lambda,\xi),p_{11}(t,\lambda,\xi))^\mathsf{T}$ and

\begin{widetext}
\bb\label{eq:Wcs}
{\cal W}(\lambda,\xi) = \left(\begin{array}{cccc}
     -W_{0}^+ &  W_{\text{H},0}^-e^{\xi \epsilon_\text{h}} & W_{\text{L},0}^-e^{-\lambda}+W_{\text{R},0}^- & 0\\
          W_{\text{H},0}^+e^{-\xi\epsilon_\text{h}} & - W_{\text{W},1}^+ - W_{\text{H},0}^-  & 0 & W_{\text{L},1}^-e^{-\lambda}+W_{\text{R},1}^-\\
    W_{\text{L},0}^+e^{\lambda}+W_{\text{R},0}^+  & 0 & -W_{\text{W},0}^- - W_{\text{h},1}^+  & W_{\text{H},1}^-e^{\xi\epsilon_\text{h}}\\
     0 & W_{\text{L},1}^{+}e^{\lambda}+W_{\text{R},1}^{+} & W_{\text{H},1}^+e^{-\xi\epsilon_\text{h}} & - W_{1}^-
\end{array}\right)
\ee
\end{widetext}
is the modified rate matrix. Counting fields involving $\lambda$ and $\xi$ have been inserted to keep track of the number of electrons coming from the left reservoir, and the energy quanta provided by the hot reservoir. 

\begin{figure}[h]
\centering
\includegraphics[width=0.49\textwidth]{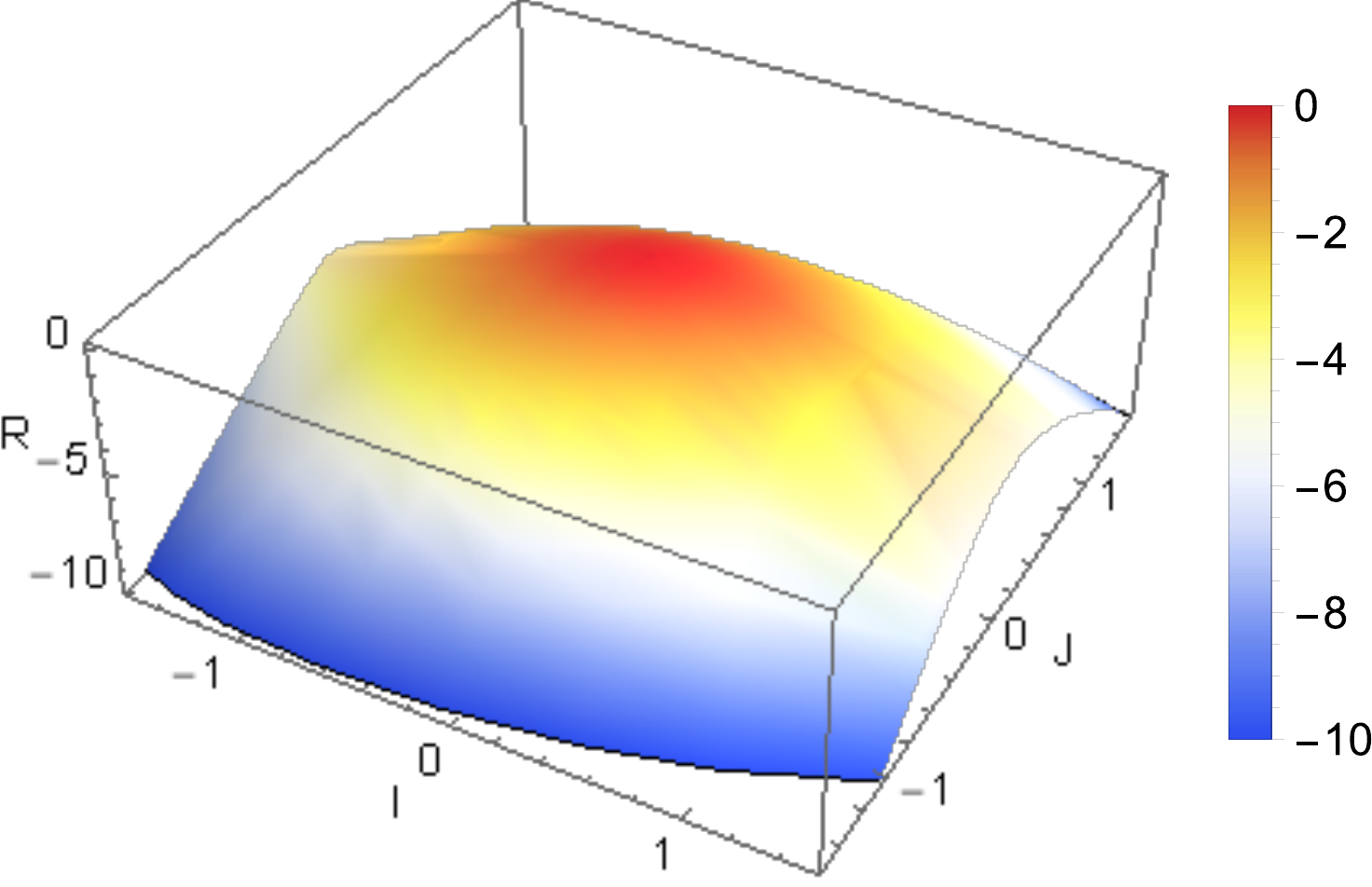}
\caption{Large deviation function as a function of the time-averaged particle current $I$ into reservoir $L$ and the time-averaged heat current $J$ from the hot reservoir along a single long trajectory. Parameters: $\epsilon_\text{w}= \epsilon_\text{h}=0$, $\Delta\mu = 0.25\Gamma$, $U = 5\Gamma$, $T_\text{H} = 2 T_\text{W} = 10\Gamma/k_B$. }\label{f:RIJ}
\end{figure}

Then, we compute the long-time cumulant generating function $S(\lambda,\xi) = \lim_{t\to\infty} \tfrac{1}{t}\ln G(t,\lambda,\xi)$ which is simply given by the dominant eigenvalue of the modified rate matrix \cite{Esposito09} (i.e. the eigenvalue with the largest real part). Finally, the long deviation function $R(I,J)$ is connected to the latter via a Legendre transform:
\bb
R(I,J) = S(\lambda,\xi) - \lambda I -\xi J,
\ee
with $I = -\partial_\lambda S(\lambda,\xi)$ and $J =  -\partial_\xi S(\lambda,\xi)$. From Eq.~\eqref{eq:PIJ}, it is clear that the maximum of $R(I,J)$ will characterize the most likely value taken by the currents along a single trajectory of long duration $t\gg \Gamma^{-1}$. The function $R(I,J)$ is plotted in Fig.~\ref{f:RIJ}. We find numerically that the peak $R(I^*,J^*)=0$ is reached for $I^*/\Gamma \simeq -\numprint{2.5E-3}$ and $J^*/\Gamma U =\numprint{1.6E-2}$ which matches the average steady-state values $\bar{I}_\text{L}$ and $\bar{J}_\text{H}$ obtained from the master equation analysis for the same parameters.

\section{Stochastic Cycles}\label{app:stoch}

A trajectory $\gamma$ can always be recast as a series of cycles $c_k \in {\cal C}$ and a remainder $\tilde\gamma$, i.e. a part that does not complete any cycle in ${\cal C}$. Interestingly, the power generated by the engine and the entropy produced along trajectory $\gamma$ of duration $\tau$ can be recast highlighting the cycle contribution \cite{Seifert12}:
\bb
 {\cal P}(\gamma) &=&\sum_{{\cal C}\in\text{cycles}} \!\!  \frac{N_{\cal C}(\gamma)}{\tau}\Delta \mu \delta n_\text{L}({\cal C})+ \frac{\Delta \mu \delta n_\text{L}(\tilde\gamma)}{\tau}\;\;\;\;\\
\dot\sigma(\gamma)&=& \sum_{{\cal C}\in\text{cycles}}  \frac{N_{\cal C}(\gamma)}{\tau} \delta\sigma({\cal C})+\frac{\delta\sigma(\tilde\gamma)}{\tau}.
\ee
Here, we have introduced the number of times $N_{\cal C}(\gamma)$, with which the cycle ${\cal C}$ occurred during trajectory $\gamma$, as well as the entropy produced $\sigma(\tilde\gamma)$ and the number of electrons transferred to the left lead $\delta n_\text{L}(\tilde\gamma)$ during the non-cyclic part of the trajectory $\tilde\gamma$. The entropy produced during cycle ${\cal C}$ is defined in Eq.~\eqref{eq:dsigC}.

Similarly, the average power and entropy production can be expressed as:
\bb
{\cal P} &=&\sum_{{\cal C}\in\text{cycles}} r_{\cal C} \Delta \mu\delta n_\text{L}({\cal C})+ \frac{\Delta\mu \delta n_\text{L}(\tilde\gamma)}{\tau},\\
\dot\sigma&=& \sum_{{\cal C}\in\text{cycles}} r_{\cal C} \delta \sigma({\cal C})+\frac{\delta \sigma(\tilde\gamma)}{\tau},
\ee
where $r_\C = \langle \! \langle N_\C(\gamma)/\tau \rangle\!\rangle \geq 0$ is the rate of occurrence of cycle $\C$ (and $\langle\!\langle \cdots \rangle\!\rangle$ is the ensemble average over trajectories). Due to the expression of the transition rates involved in the cycles, the rate of occurrence of cycle ${\cal C}$ and its reverse $\bar{\cal C}$  are related via:
\bb
\frac{r_{\cal C}}{r_{\bar{\cal C}}} = e^{\delta\sigma({\cal C})},
\ee
such that we have
\bb
{\cal P} &=&\sum_{{\cal C}\in\text{cycles}^+} r_{\cal C}(1-e^{-\delta\sigma({\cal C})}) \Delta\mu\delta n_\text{L}({\cal C})\nonumber\\
&&\quad\quad\quad\quad+ \moy{\frac{\Delta\mu\delta n_\text{L}(\tilde\gamma)}{\tau}},\label{C:P}\\
\dot\sigma&=& \sum_{{\cal C}\in\text{cycles}^+}j_c(1-e^{-\sigma(c)}) \delta\sigma(c)+\moy{\frac{\sigma(\tilde\gamma)}{\tau}},\quad\label{C:sig}
\ee
with the sums running over the direct cycles only, defined such that entropy production $\delta\sigma({\cal C}) = - \delta\sigma(\bar{\cal C}) \geq 0$.  In Eqs.~\eqref{C:P}-\eqref{C:sig}, we have used the double angle brackets to denote the stochastic average over the final portion of the trajectories, not part of any cycle.
The non-cyclic part of the trajectory $\tilde\gamma$ generates a work output typically bounded by $2\Delta \mu$ and entropy production $\vert\sigma (\tilde\gamma) \vert < 2(\beta_\text{W}-\beta_\text{H})U+2\Delta\mu$, such that the terms $\moy{\frac{\Delta\mu\delta n_\text{L}(\tilde\gamma)}{\tau}}$, $\moy{\frac{\sigma(\tilde\gamma)}{\tau}}$ go to zero when $\tau\to\infty$. On the other hand, for sufficiently long trajectories, the work and the entropy produced scale as $t$ (the engine reaches a non-equilibrium steady state characterized by a constant current to reservoir L). This is only possible if the contribution of the cycles does not vanish for long times.

\section{Analytical expression for cycle currents and analytical evaluation of stopping bias}\label{app:cyclecurr}

\begin{figure}[h!]
  \includegraphics[width=0.49\textwidth]{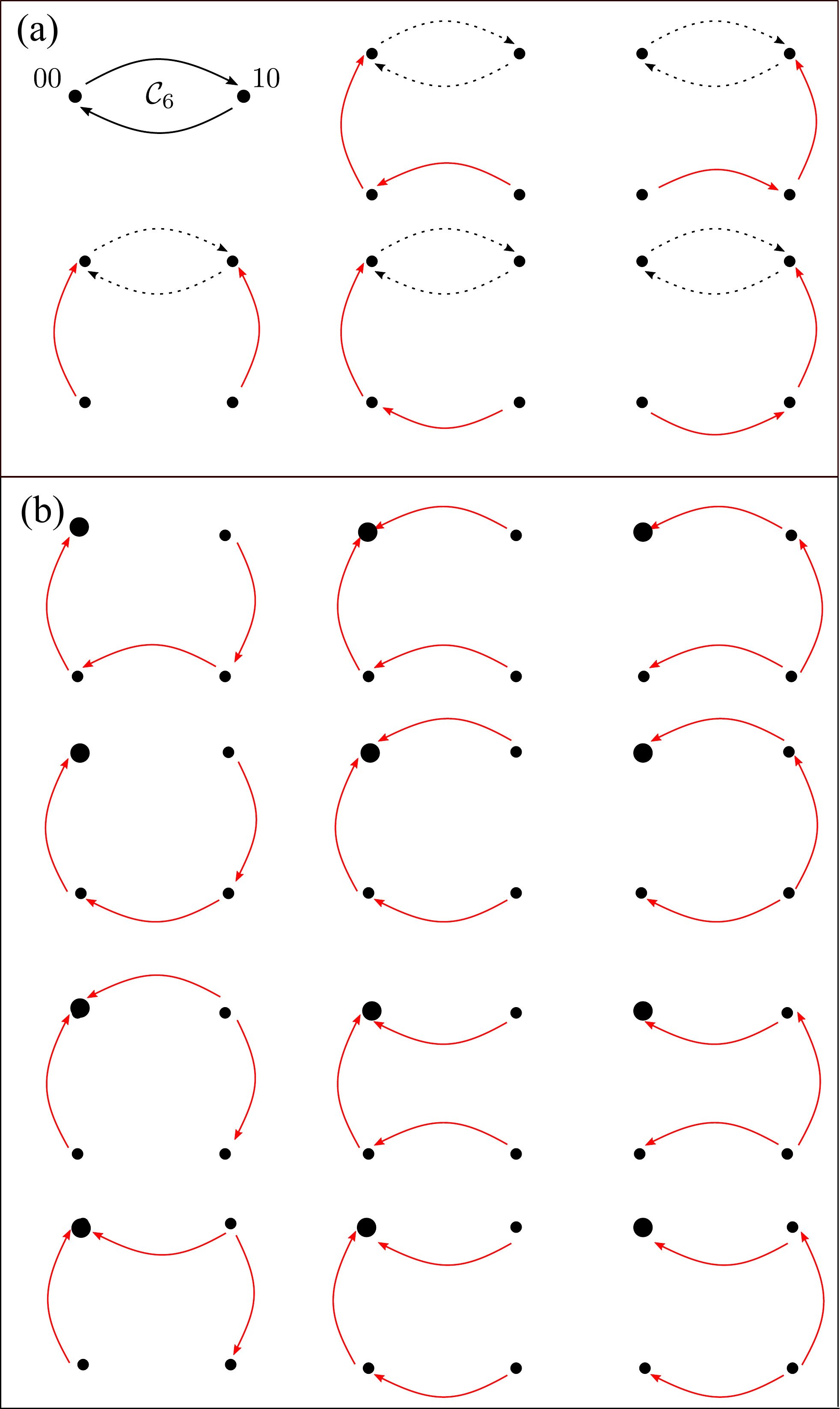}
\caption{(a) Diagrams contributing to the term $\Sigma_{\C_6}$, associated with cycle ${\C_6}$ represented in solid black. The four dot represent the four two-dot states and the arrows are drawn from the possible transition in the Markov chain describing the system dynamics (see Fig.~\ref{f:Markov}). The contribution of each diagram is the product of the rates associated with the red arrows. The cycle ${\cal C}_{6}$ is recalled in each diagram with dashed black arrows to show how the red arrows converge towards one of the states involved in the cycle. (b) Sample of the diagrams contributing to the term $\Sigma$. The contribution of each diagram is the product of the rates associated with the red arrows. Here only the diagrams involving sequences of arrows converging to state $00$ (represented by a larger dot) are represented. The other diagrams converging towards states $ij\neq 00$ can be deduced by symmetry.}\label{f:DecompC6}
\end{figure}

The average oriented cycle currents (or occurrence rates of oriented cycles) $r_\C$ can be expressed analytically from the tools of graph theory \cite{HillBook}, according to:
\bb
r_\C = \frac{\Pi_\C \Sigma_\C}{\Sigma}
\ee
Here, $\Pi_\C$ is the product of the rates of the transitions involved in the cycle. For a cycle involving not all the states of the graph, $\Sigma_\C$ is the sum of products of transition rates corresponding to sub-graphs which (i) contain the maximum number of edges without creating a new cycle when they are added to cycle $\C$ and (ii) with arrows or sequences of arrows converging towards one state involved in the cycle. For a cycle involving all the states of the graph,  $\Sigma_\C =1$. Finally, $\Sigma$ is the sum of products of transition rates corresponding to sub-graphs which contain (i) the maximum number of edges without creating any cycle, and (ii) contains arrow or sequences of converging towards \emph{any of the four states} of the graph. These rules are illustrated graphically in the case of cycle $\C_6$ in Fig.~\ref{f:DecompC6}. We focus on the cycles contributing to the particle transport, i.e. ${\cal C}_1$, ${\cal C}_4$, ${\cal C}_6$ and the spurious sub-cycle involving transitions $L_+R_-$ while $n_\text{h} = 1$, that we denote ${\cal C}^*$.
We find:

\bb
\Pi_{\C_1} &=& W_{\text{L}0}^+ W_{\text{H}1}^+ W_{\text{R}1}^- W_{\text{H}0}^-,\\
\Pi_{\C_4} &=& W_{\text{R}0}^+ W_{\text{H}1}^+ W_{\text{L}1}^- W_{\text{H}0}^-,\\
\Pi_{\C_6} &=& W_{\text{L}0}^+ W_{\text{R}0}^-,\\
\Pi_{\C^*} &=& W_{\text{L}1}^+ W_{\text{R}1}^-,\
\ee
and
\bb
\Sigma_{\C_1} &=& 1,\\
\Sigma_{\C_4} &=& 1,\\
\Sigma_{\C_6} &=& W_{\text{H}0}^- W_{\text{H}1}^- + W_{\text{W}1}^- W_{\text{H}0}^- + W_{\text{W}1}^+ W_{\text{H}1}^-,\\
\Sigma_{\C^*} &=& W_{\text{H}0}^+ W_{\text{H}1}^+ + W_{\text{W}0}^- W_{\text{H}0}^+ + W_{\text{W}0}^+ W_{\text{H}1}^+.
\ee

Finally, the term $\Sigma$ contains $48$ terms, $12$ of which are graphically represented in Fig~\ref{f:DecompC6}(b). As this term does not depend on the cycle $\C$, we do not need to evaluate it to compute the ratio of cycle average currents. We therefore have:
\bb
\frac{r_{\C_4}}{r_{\C_6}} = \frac{\Pi_{\C_4}}{\Pi_{\C_6}\Sigma_{\C_6}}.
\ee
Similarly we can take into account the contribution of cycle $1$ and $5$ using:
\bb
\frac{r_{\C_4}}{r_{\C_1}+r_{\C_5}+r_{\C_6}} = \frac{\Pi_{\C_4}}{\Pi_{\C_1}  +\Pi_{\C_6}\Sigma_{\C_6}+\Pi_{\C^*}\Sigma_{\C^*}}.
\ee

The average intensity can be computed as:
\bb
\bar{I}_\text{L}  &=& r_{\C_4} - r_{\Cb_4} \nonumber\\
&&- (r_{\C_1} - r_{\Cb_1}+r_{\C_6}-r_{\Cb_6}+r_{\C^*} - r_{\Cb^*})\nonumber\\
&=& r_{\C_4}(1-e^{-\delta\sigma(\C_4)}) - (r_{\C_1}  +r_{\C_6}+r_{\C^*})(1-e^{-\delta\sigma(\C_6)}).\nonumber\\
\ee
Solving $\frac{r_{\C_4}}{r_{\C_1}+r_{\C_6}+r_{\C^*}} = \frac{1-e^{-\delta\sigma(\C_6)}}{1-e^{-\delta\sigma(\C_4)}}$ for the value of $\Delta\mu$ yields the stopping bias. For the parameters of Fig.~\ref{num_cycles}, we find $\Delta \mu_\text{stop} = 0.62\Gamma$.

%


\end{document}